\documentclass[usenatbib,letterpaper]{mn2e}
\usepackage[letterpaper,totalwidth=500pt,totalheight=680pt,layoutvoffset=0.5cm]{geometry}

\usepackage{graphicx}
\usepackage{epsfig}
\usepackage{epsf}
\usepackage{amsmath}
\usepackage{amssymb}
\usepackage{stfloats}
\usepackage{caption}
\usepackage{subcaption}
\usepackage{lscape}
\usepackage{deluxetable}
\usepackage{epstopdf}
\usepackage[hyphens]{url}

\usepackage{booktabs}
\usepackage{threeparttable}

\newcommand{\spitzer}{\mbox {\em Spitzer}}
\newcommand{\um}{$\mu$m}
\newcommand\ion[2]{#1$\;${\small\rmfamily\MakeUppercase{\romannumeral#2}}\relax}
\newcommand{\hii}{\mbox{\ion{H}{2}~}}
\renewcommand{\deg}{$^\circ$ }
\newcommand{\degree}{$^{\circ}$ }

\newcommand{\kms}{km~s$^{-1}$}

\title[EROs in the Carina Nebula]{Extended Red Objects and Stellar
  Wind Bow Shocks in the Carina Nebula}

\author[Sexton et al.]{Remington O.\ Sexton$^{1,2}$\thanks{Email:
    remington.sexton@email.ucr.edu}, Matthew S.\ Povich$^1$, Nathan
  Smith$^3$, Brian L.\ Babler$^4$, \and Marilyn R. Meade$^4$, \& Alexander L. Rudolph$^1$ \\
  $^1$Department of Physics \& Astronomy, California State Polytechnic
  University, 3801 West Temple Avenue, Pomona, CA 91768, USA \\
  $^2$Department of Physics and Astronomy, University
  of California, Riverside, 900 University Avenue, Riverside, CA
  92521, USA \\
  $^3$Steward Observatory, University of Arizona, 933 North Cherry
  Avenue,
  Tucson, AZ 85721, USA \\
  $^4$Department of Astronomy, University of Wisconsin, 475 North
  Charter Street, Madison, WI 53706, USA}

\begin{document}

\pagerange{\pageref{firstpage}--\pageref{lastpage}} \pubyear{2014}
\maketitle
\label{firstpage}

\begin{abstract}
  We report the results of infrared photometry on 39 extended red
  objects (EROs) in the Carina Nebula, observed with the {\em Spitzer
    Space Telescope}.  Most EROs are identified by bright, extended
  8.0~\um\ emission, which ranges from 10\arcsec\ to 40\arcsec\ in
  size, but our sample also includes 4 EROs identified by extended
  24~\um\ emission.
  Of particular interest are nine EROs associated with late O or early
  B-type stars and characterized by arc-shaped morphology, suggesting
  dusty, stellar-wind bow shocks.
  These objects are preferentially oriented towards
  the central regions of the Carina Nebula, suggesting that these bow
  shocks are generally produced by the interactions of OB winds with
  the bulk expansion of the \hii
  region 
  rather than high proper motion.  We identify
  preferred regions of mid-infrared color space occupied by our bow
  shock candidates, which also contain bow shock candidates in M17 and
  RCW 49 but are well-separated from polycyclic aromatic hydrocarbon
  emission or circumstellar discs. Color cuts identify an additional 12
  marginally-resolved bow shock candidates, 10 of
  which are also associated with known late O or early B stars.  \hii
  region expansion velocities derived from bow shock candidate
  standoff distances are ${\sim}10$~km~s$^{-1}$, assuming typical \hii
  region gas densities, comparable to expansion velocities derived
  from bow shocks in M17 and RCW 49.  One candidate bow shock provides
  direct evidence of physical interaction between the massive stellar
  winds originating in the Trumpler 15 and Trumpler 14 clusters,
  supporting the conclusion that both clusters are at similar
  heliocentric distances.
\end{abstract}


\section{Introduction}

Massive stars are the cosmic engines that drive the turbulent
evolution of their parent nebulae, creating observable feedback
effects such as stellar-wind bow shocks.  Wind velocities typical of
massive late-O to early-B type stars can range between
900--3000~km~s$^{-1}$ \citep{Fullerton}, exceeding the sound speed in
their ambient medium \citep{Castor} and forming ionized shock fronts.
If the driving star itself has a supersonic velocity relative to the
surrounding medium, ambient gas and dust can be swept around the star,
forming a stellar-wind bow shock \citep{vanBuren2}.  {\em Spitzer
  Space Telescope} observations of the RCW 49 and M17 \hii regions
made as part of the Galactic Legacy Infrared Mid-Plane Survey
Extraordinaire \citep[GLIMPSE;][]{GLIMPSE2} revealed such stellar-wind
bow shocks forming around late O to early B-type stars
\citep[][hereafter P08]{Povich}, and similar dusty shock structures
have been found in other massive star-forming regions
\citep{Kobulnicky,Gvaramadze}.  \citet[][hereafter S10]{Smith}
identified 8 ``extended red objects'' (EROs) in {\em Spitzer} images
of the South Pillars region of the Great Nebula in Carina,
characterized by bright, extended emission in the 8.0~$\mu$m bandpass
of the Infrared Array Camera \citep[IRAC;][]{Fazio}, and suggested
that these objects are dusty bow shocks driven by late O and early B
stars, similar to those found by P08.


Located in the Sagittarius--Carina spiral arm at a distance of 2.3~kpc
\citep{Allen,Smith2002,Smith2006a}, the Carina Nebula is one of the
few extreme massive star-forming regions close enough for detailed
observation and unobscured by substantial reddening.  It contains a
large and varied population of high-mass stars, including 3 WNH stars,
and over 70 O-type stars, some of which among the earliest ever
discovered \citep{Walborn1}.  The central region of the Carina Nebula
contains two massive clusters, Trumpler (Tr) 16 and Tr 14, which are
the two largest contributors of stellar luminosity to the nebula
\citep{Smith2006b}.  Tr 16 itself contains 43 known O-type stars, one
of which is the luminous blue variable $\eta$ Car \citep{Smith2006b}.
The combined ionizing luminosity, radiation pressure, and stellar
winds from the other massive stars in Tr 16, Tr 14, and other
sub-clusters throughout the Carina Nebula
\citep[S10;][]{CCCP_clustering} provide the feedback that drives the
physical evolution of the nebula
\citep{Smith2000}. 

The formation of bow shock structures around stars in and near \hii
regions gives insight into how massive star feedback shapes the local
star-forming environment.
Bow shock structures can be produced by the interaction between
stellar winds from their driving stars and large-scale gas flows
produced by the global expansion of \hii regions and photoevaporation
at the interfaces between \hii regions and surrounding cold, molecular
gas (P08, S10). Bow shocks also serve as a means for identifying
candidate ``runaway'' OB stars escaping from their natal \hii regions
at high velocities \citep{RunawaysI, Kobulnicky, RunawaysII}.

Extending the work of S10, we identify 39 EROs from {\em Spitzer}/IRAC
imaging of the full spatial extent of the Carina Nebula. The rest of
this paper is organized as follows: In Section 2 we describe our ERO
selection criteria, measurements of their geometries, and infrared
(IR) photometry. In Section 3 we report results of these measurements,
including the morphological classification of 9 EROs as candidate bow
shocks and the classification of an additional 12 EROs as candidate
bow shocks based on analysis of their mid-IR colors.
In Section 4 we use measured (projected) standoff distances, empirical
mass-loss rates and stellar wind velocities to estimate the velocity
of the ambient interstellar medium (ISM) relative to the driving
stars.  We also discuss EROs that are not evident bow shock
candidates, including two anomalous EROs, both of which exhibit
resolved bow shock morphology, but fail on the criterion of color. Our
conclusions are summarized in Section 5.


\section{Observations and Data Analysis}

\footnotetext[1]{See \url{http://irsa.ipac.caltech.edu/data/SPITZER/GLIMPSE/doc/velacar\_dataprod\_v1.0.pdf}}
\footnotetext[2]{Available at \url{http://www.astro.wisc.edu/glimpse/glimpse\_photometry\_v1.0.pdf}}

The \emph{Spitzer} Vela-Carina survey \footnotemark[1] 
(S. R. Majewski, PI) observed the entire Carina Nebula using all four
IRAC bandpasses, centreed at 3.6, 4.5, 5.8, and 8.0 $\mu$m, with a
resolution of ${\sim}2\arcsec$ \citep{Majewski,Povich3}.  Observations
of the Carina Nebula with the Multiband Imaging Photometer for {\em
  Spitzer} \citep[MIPS;][]{Rieke} carried out during July 2007 as part
of program GO-30848 (MIPSCAR; N. Smith, PI) provided 24.0 $\mu$m
imaging with approximately 6$''$ resolution \citep{Povich3}.
The sensitivity of the IRAC and MIPS images to EROs was dramatically
reduced in the central regions of the Carina nebula by the extreme
mid-IR brightness of $\eta$ Car and source confusion in the dense
clusters Tr 14 and Tr 16. $\eta$ Car itself was intentionally avoided
by the MIPSCAR observation.  The individual IRAC and MIPS frames were
combined into large mosaics (1.2\arcsec\ pixel scale) using the
GLIMPSE data reduction pipeline \citep{GLIMPSE1, GLIMPSE2}.

``Residual images,'' or mosaics with the point sources removed, are an
advanced data product produced by the GLIMPSE pipeline and used in
this study, as they facilitate photometry of extended sources such as
EROs. The residual images (images with point sources removed) were
produced by performing Point Spread Function fitting using a modified
version \footnotemark[2]
(Babler 2006) of DAOPHOT \citep{Stetson} on individual IRAC
frames. The residual image tiles are mosaics from individual frame
residuals.  Thus if a source is extracted in some but not all frames
it will show up in these images as a source. Sources may not be
extracted for a variety of reasons, mainly due to cosmic ray
contamination, saturation/non-linearity limits and along frame edges.

The data used by S10 consisted of a earlier and deeper set of IRAC
observations limited to the South Pillars and south-western region of
the nebula, avoiding the bright central regions surrounding Tr 16 and
Tr 14 as well as the fields to the north.  Utilizing the wider-field,
Vela-Carina survey data allowed us to search the nebula in its
entirety and enlarge our ERO candidate sample.

As part of an observing campaign targeting known and candidate OB
stars in the Carina Nebula (NOAO Proposal \#2013A-0181; M. S. Povich,
PI), on 18 March 2013 we obtained spectra at 3975--4225 \AA\ and
4250--4500 \AA\ for 15 candidate ERO driving stars, including 4
lacking previous classifications, using the 2dF/AAOmega spectrgraph on
the Australian Astronomical Telescope (AAT).  O stars and early B
(B0-B2) stars were classified using the criteria and digital atlas of
\citet{WF90}. Relative strengths of \ion{He}{2} 4200 and
\ion{He}{1+II} 4026 were used most often to determine the spectral
class for O stars.  B stars were classified primarily using the
strengths of \ion{He}{1} lines (strongest at B2) the
\ion{C}{3}+\ion{O}{2} blend near 4070, and \ion{Si}{4} 4089.
Luminosity classes I and III were assigned primarily by the strength
of \ion{Si}{4} 4089 and Balmer line absorption.

\begin{figure*}
\centering
	\begin{subfigure}[b]{0.4\textwidth}
	\includegraphics[scale=0.65]{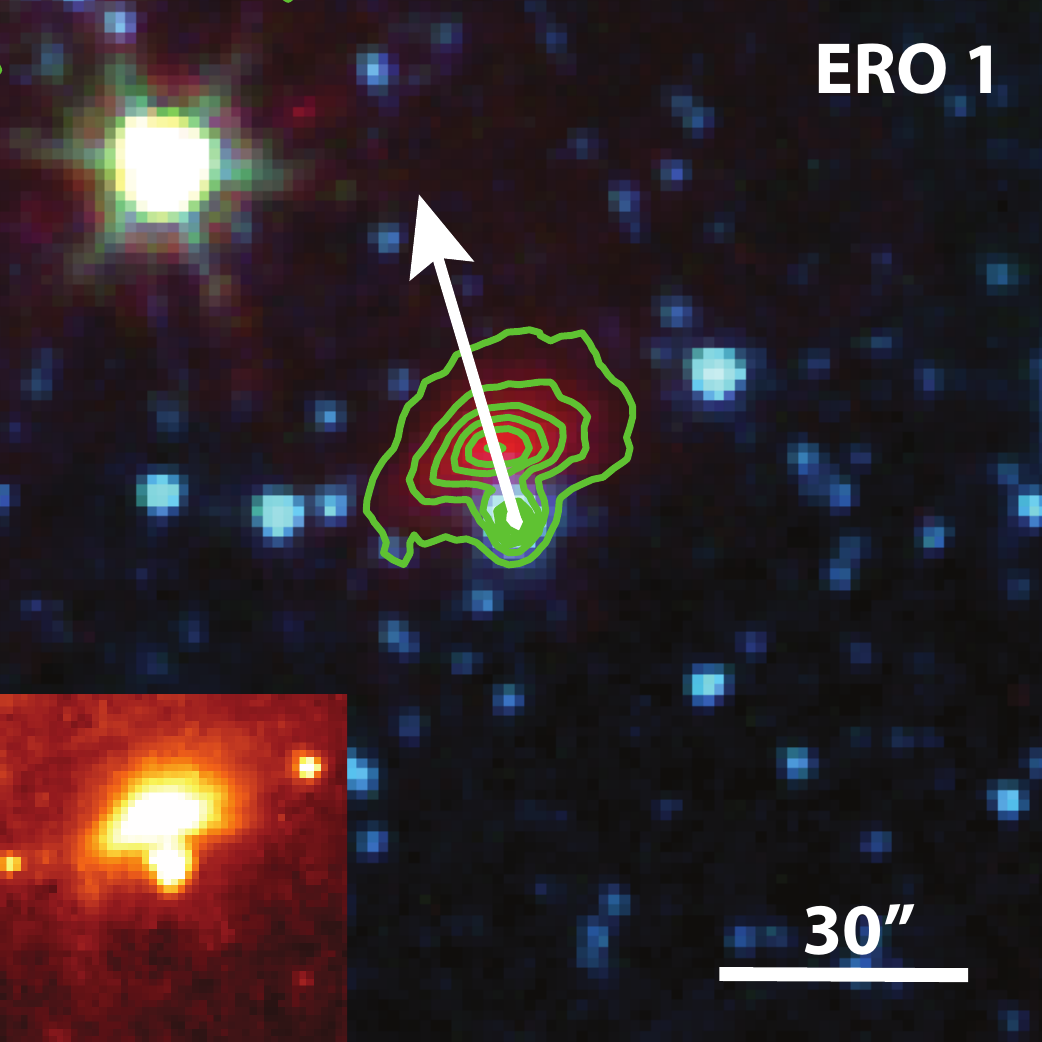}
	\end{subfigure}
	\begin{subfigure}[b]{0.4\textwidth}
	\includegraphics[scale=0.65]{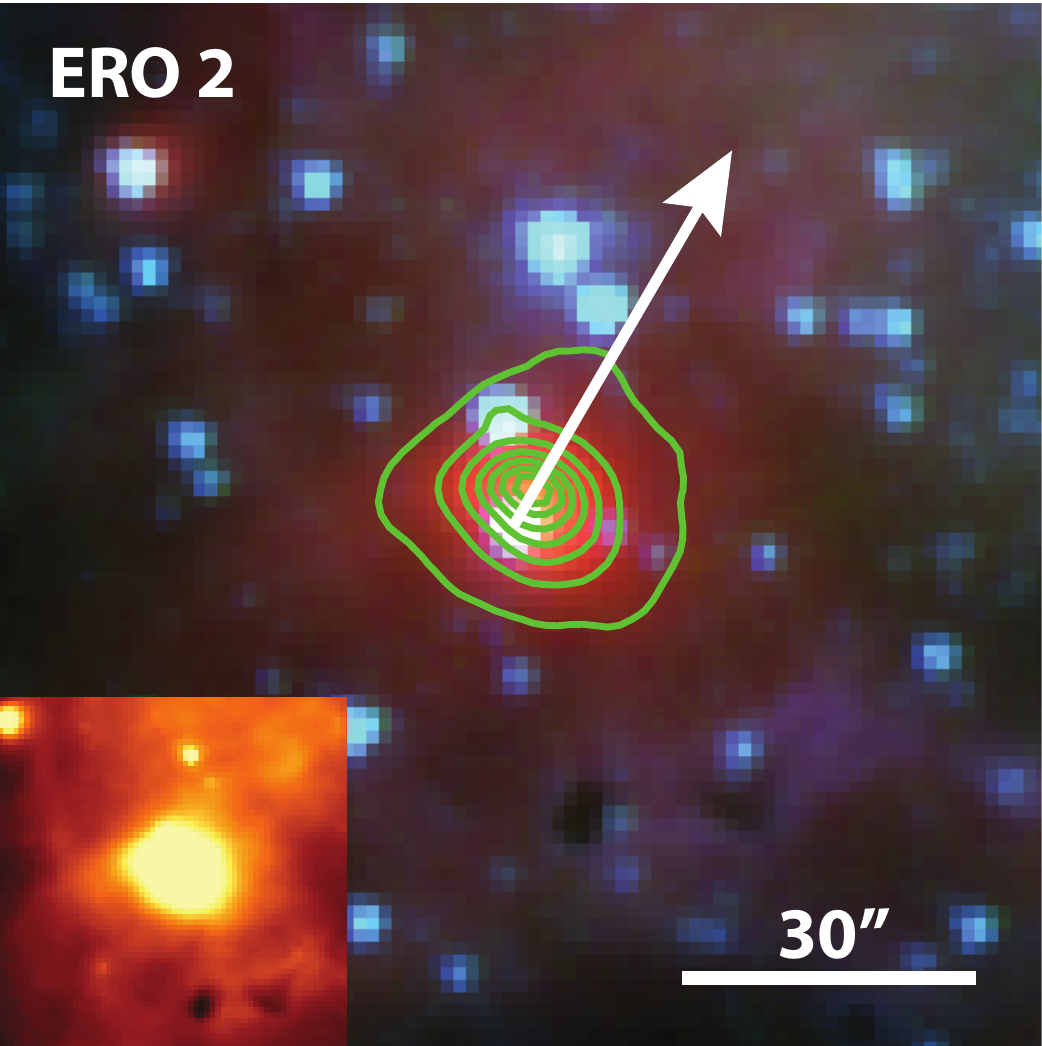}
	\end{subfigure}

	\begin{subfigure}[b]{0.4\textwidth}
	\includegraphics[scale=0.65]{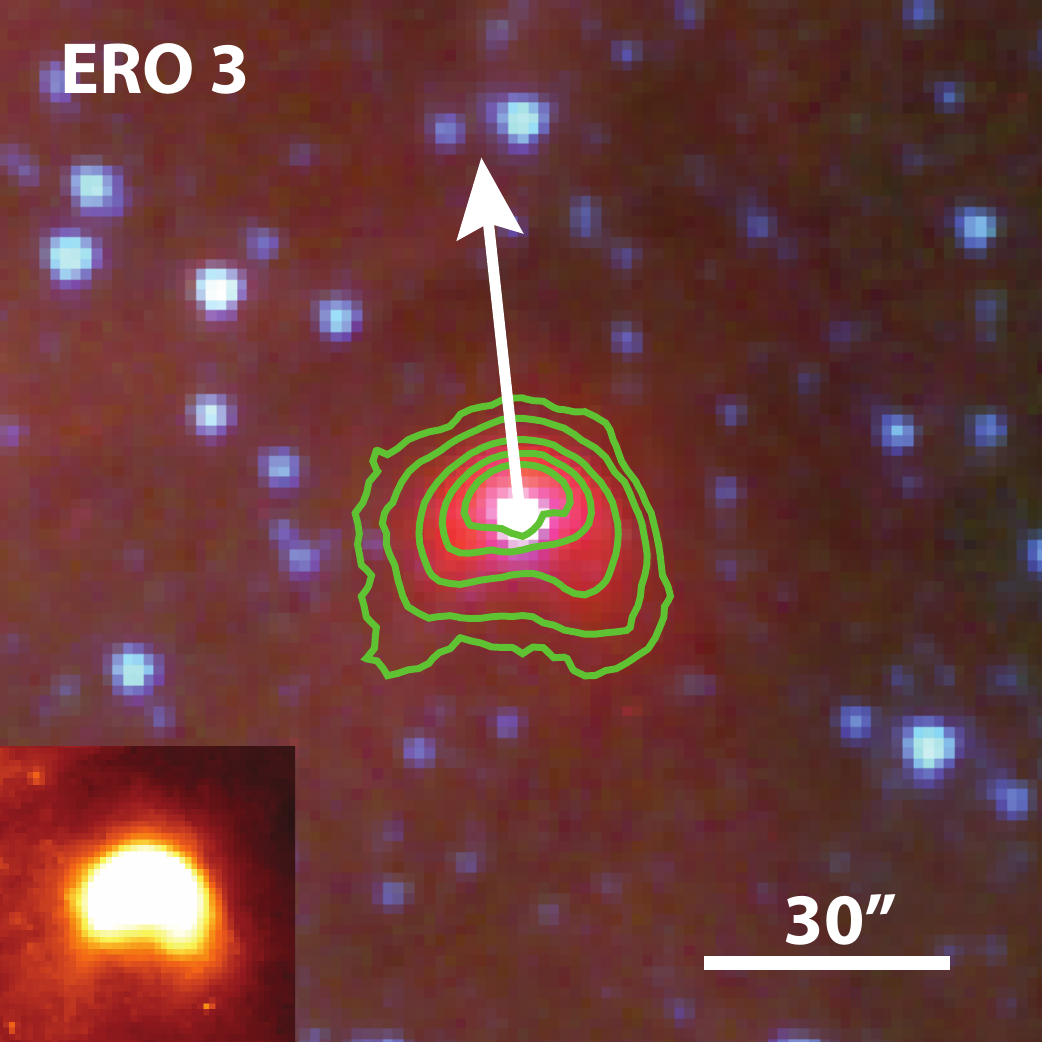}
	\end{subfigure}
	\begin{subfigure}[b]{0.4\textwidth}
	\includegraphics[scale=0.65]{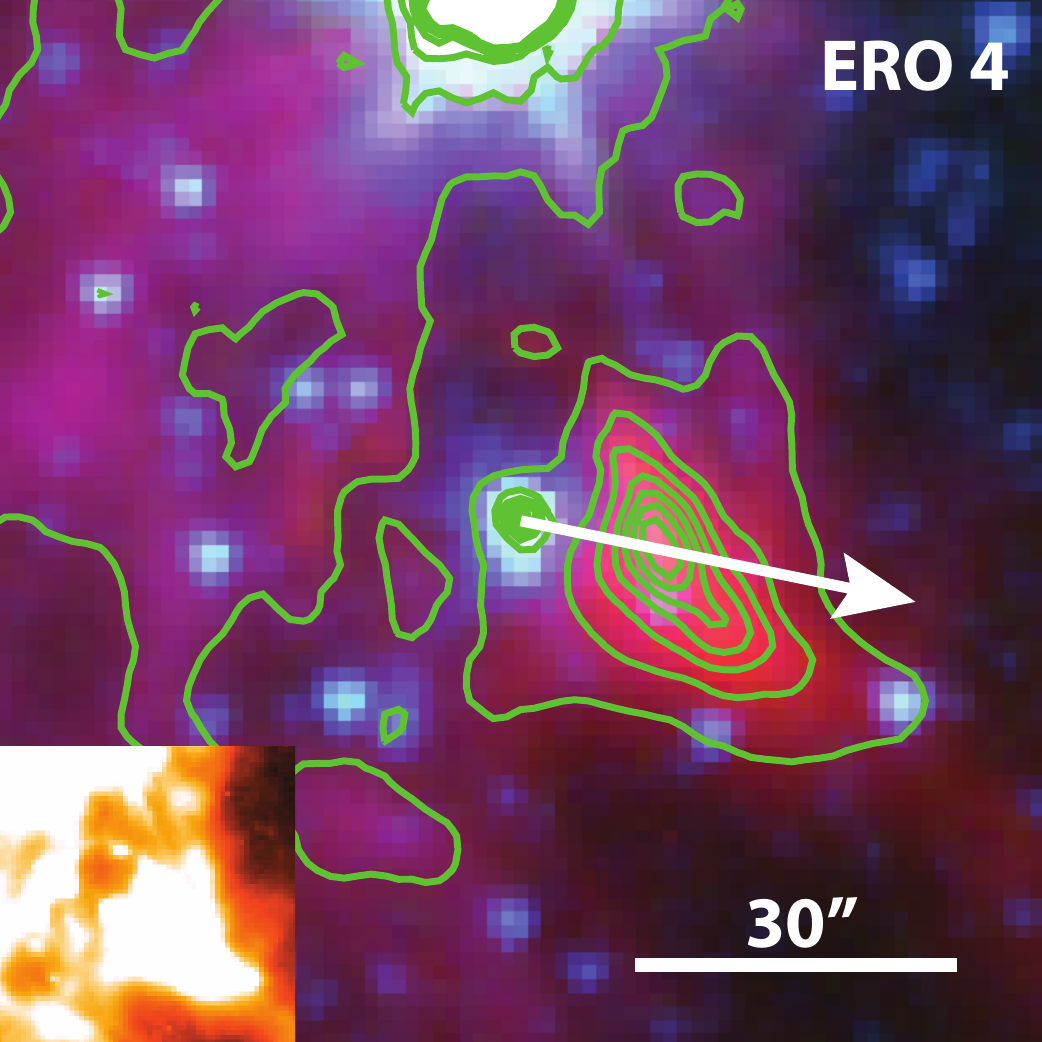}
	\end{subfigure}

	\begin{subfigure}[b]{0.4\textwidth}
	\includegraphics[scale=0.65]{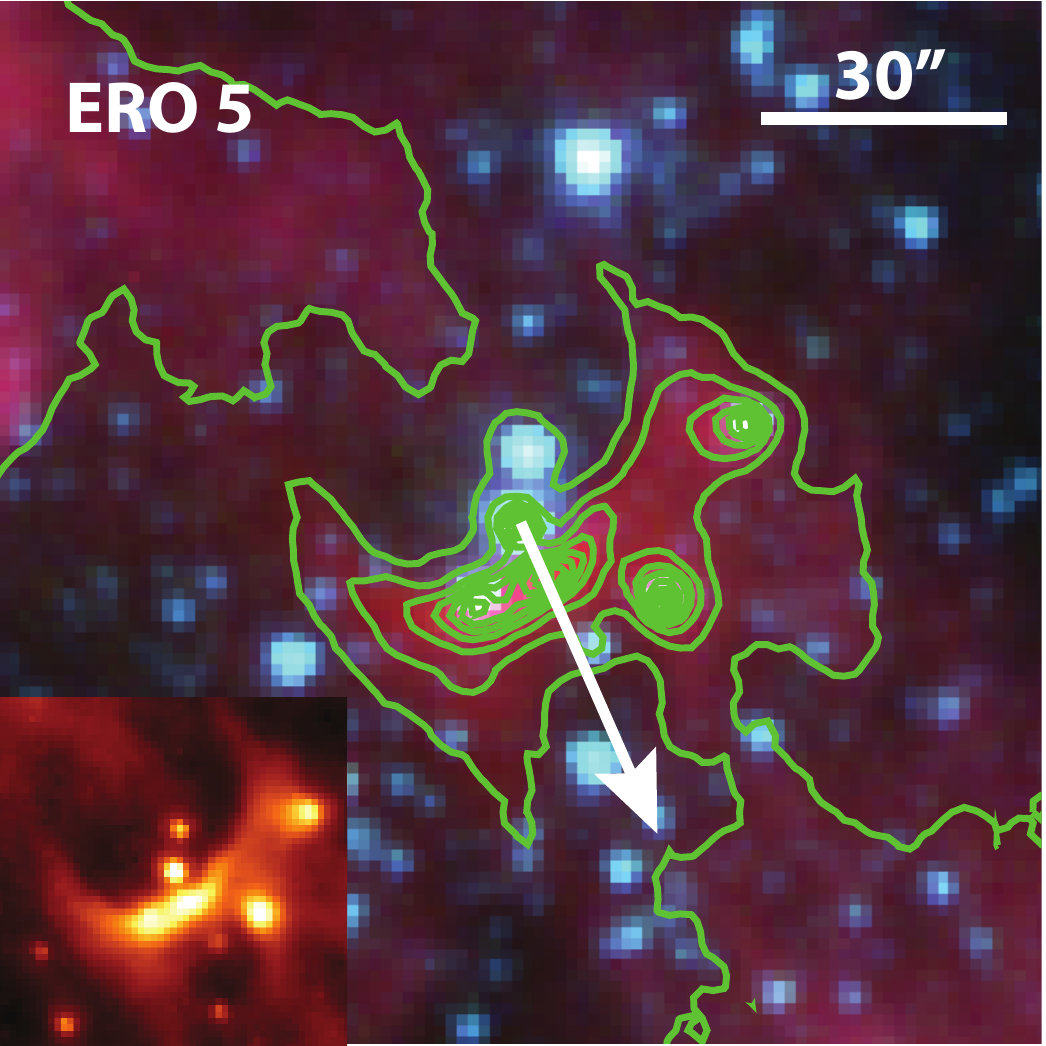}
	\end{subfigure}
	\begin{subfigure}[b]{0.4\textwidth}
	\includegraphics[scale=0.65]{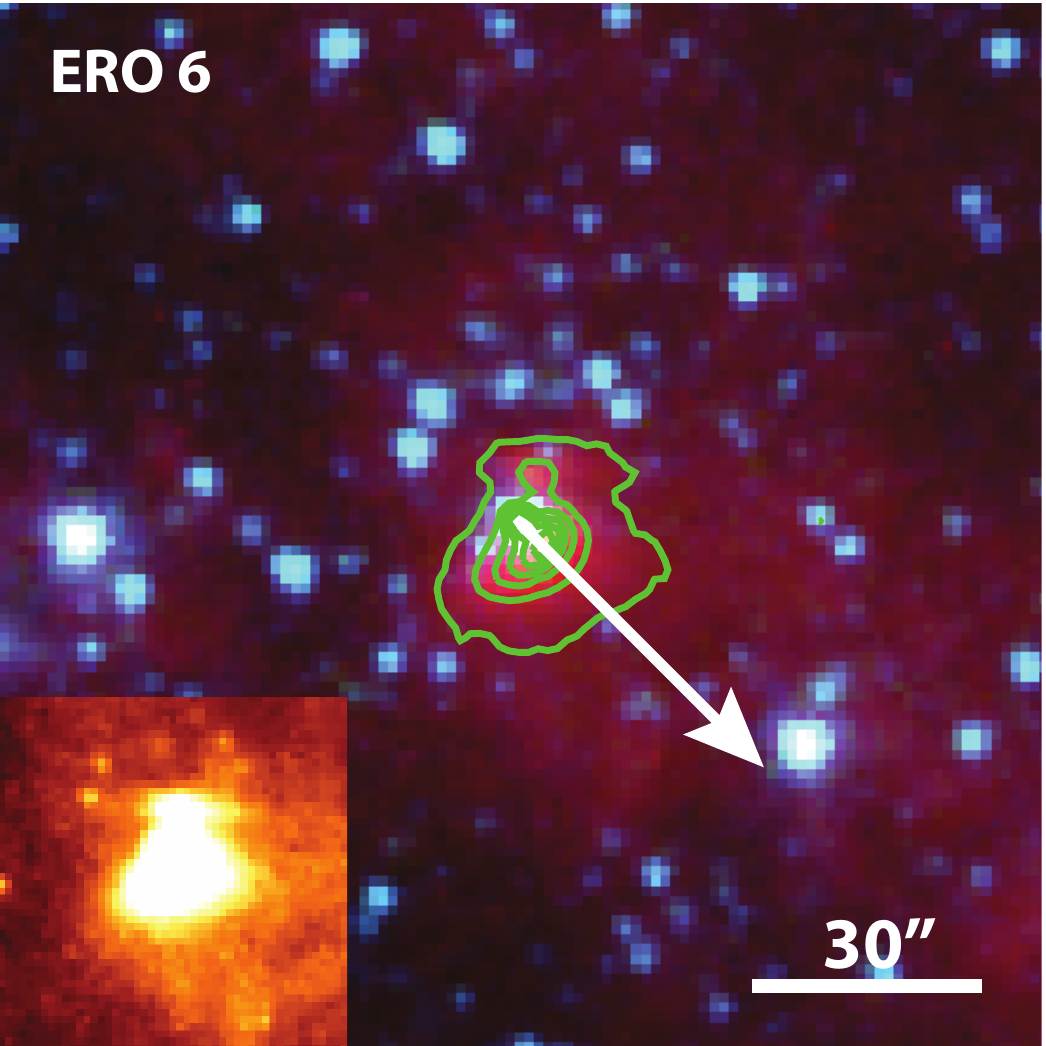}
	\end{subfigure}
\caption{Composite 3-color images of bow shock candidates from ERO selection criterion (blue = 3.6 $\mu$m, green = 4.5 $\mu$m, red = 8.0 $\mu$m).  An 8.0 $\mu$m image of each ERO is inset at the bottom left of each image.  Linearly-scaled 8.0~\um\ contours are overlaid to outline the shape of each ERO.  Arrows indicate the preferential orientation of each ERO. 
}
\label{fig:EROs}
\end{figure*}

\subsection{Identification of EROs}

ERO candidates are characterized by extended 8.0 $\mu$m emission,
ranging from 10$''$ to 40$''$ in size
(Table~\ref{tab:EROdrivingstars}).  The original S10 ERO selection
criterion only identified EROs with some apparent arc-shaped
morphology.
We widened the original S10 selection criteria to include any visible
extended emission at 8.0 $\mu$m in close proximity to or shrouding a
possible driving star, but apparently isolated from larger nebular
structures.  Our revised selection criteria, as well as the use of the
wider-field Vela--Carina mosaics, allowed us to identify 32 additional
candidate EROs beyond those originally found by S10, without bias for
ERO emission, shape, or orientation, for a total sample of 39 EROs,
including 7 from S10 (ERO S1 from S10 was excluded from our sample,
because it lacks an obvious driving star and is not sufficiently
isolated from larger nebular structures).
EROs 1--6 (Table~\ref{tab:EROdrivingstars} and Figure~\ref{fig:EROs})
exhibit clearly resolved bow shock morphologies, or asymmetrical arc
shapes offset from the driving star, similar to those seen in P08.


\begin{figure*}
\centering
	\begin{subfigure}[b]{0.4\textwidth}
	\includegraphics[scale=0.55]{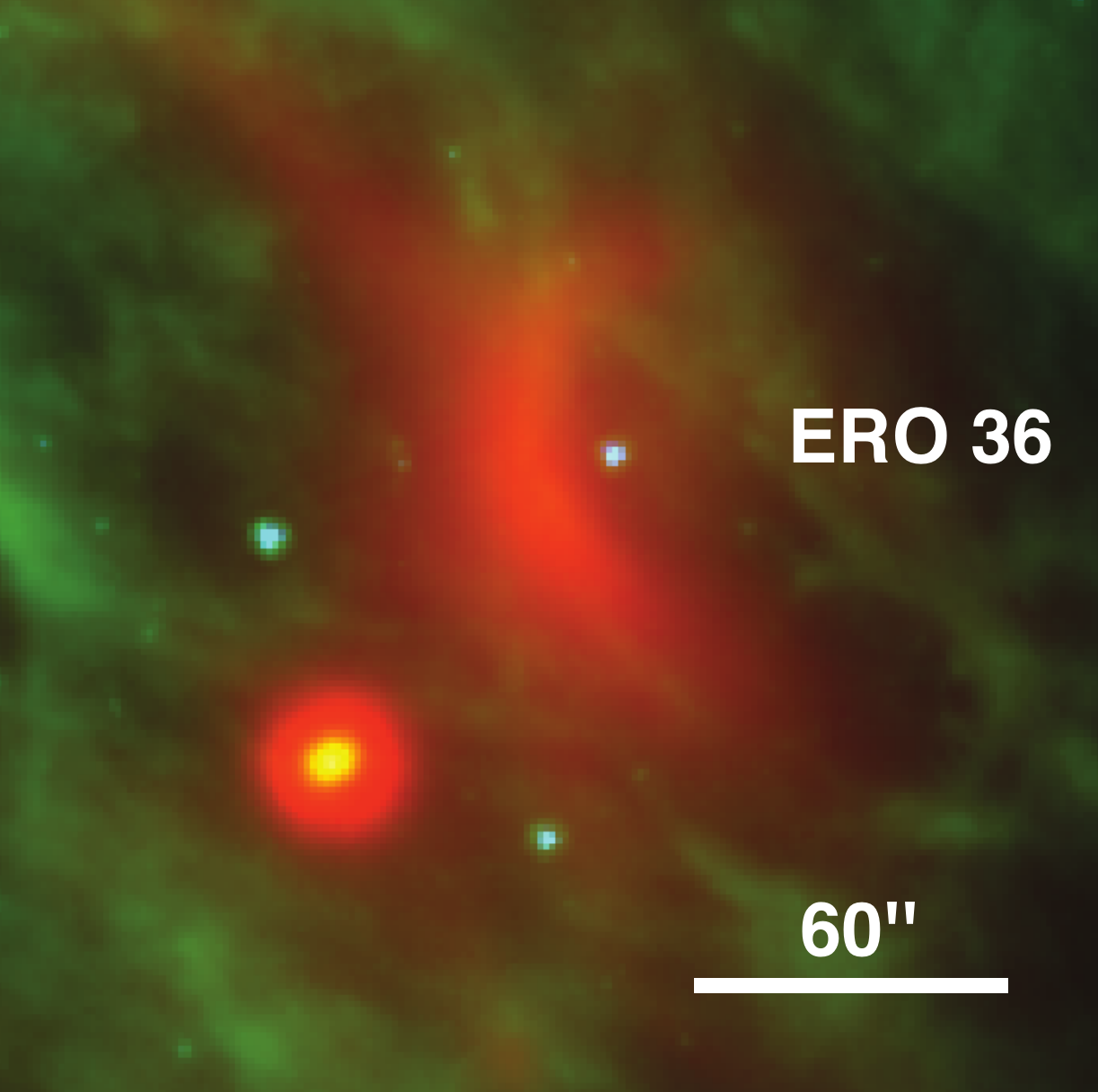}
	\end{subfigure}
	\begin{subfigure}[b]{0.4\textwidth}
	\includegraphics[scale=0.55]{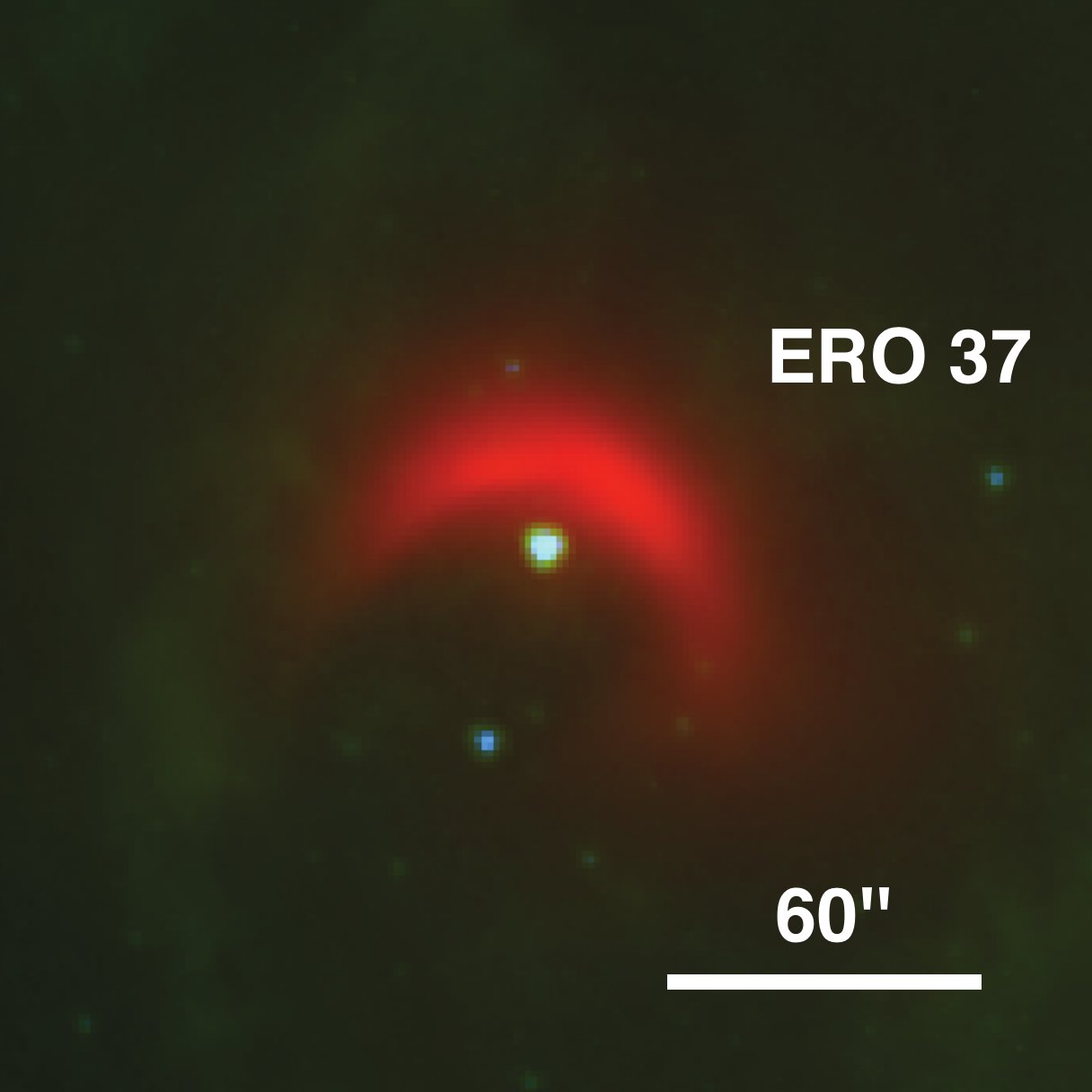}
	\end{subfigure}
\caption{Composite 3-color images of the two 24.0 $\mu$m bow shock
  candidates with the clearest arc morphologies (blue = 4.5 $\mu$m,
  green = 8.0 $\mu$m, red = 24.0
  $\mu$m). 
}
\label{fig:24candidates}
\end{figure*}


Following \citet{Kobulnicky}, we also identify any extended objects of
interest in the 24.0 $\mu$m MIPS images using analogous selection
criteria to the 8.0~\um\ EROs.  Because much of the Carina Nebula
itself produces very bright diffuse 24~\um\ emission, our search was
generally confined to the periphery of the nebula where background
surface brightness was at a minimum.  A total of four 24.0 $\mu$m
objects were found (EROs 36--39 in Table~\ref{tab:EROdrivingstars}),
three of which, EROs 36, 37, and 39, exhibit arc-shaped morphologies
(Figure~\ref{fig:24candidates}).

\begin{table*}
\begin{threeparttable}
\begin{minipage}{165mm}
\caption{\label{tab:EROdrivingstars} ERO positions, geometric properties, and probable driving stars.  EROs 1-35 are objects identified using the 8.0 $\mu$m IRAC images; EROs 1-6 are additionally morphological bow shock candidates.  EROs 36-39 were identified using the 24.0 $\mu$m MIPS images. EROs 1, 3, 4, 18, 24, 25, and 31 were previously identified by S10.  Coordinates refer to the positions of the candidate ERO driving stars.  Unless indicated otherwise, published spectral types are from \citet{Gagne}. New AAT spectral types are from our 2dF/AAOmega spectroscopic observations.}
\begin{tabular}{lllccclcc}
\toprule[1.5pt]
ERO & $\alpha_{2000}$ & $\delta_{2000}$ & Size       & Offset & Orientation 		& Star & \multicolumn{2}{c}{Spectral Type}\\
&&& (asec) & (asec) & (deg W of N) & & Published & AAT\\
\midrule
\multicolumn{8}{c}{EROs showing resolved bow shock morphology} \\ 
1     & 10:43:17.907 & -60:08:03.460 & 16.7  & 8.6 & -13 & HD 93027  & O9.5 IV & O9.5 V \\
2     & 10:44:00.904 & -59:35:46.040 & 15.3  & 4.7 & 34 & Tr 14-127 & O9 V & O9 V \\
3     & 10:44:11.116 & -60:03:21.283 & 20.9  & 3.0 & -6 & HD 305536  & O9.5 V & \\
4     & 10:44:36.204 & -60:05:29.074 & 32.9  & 12.1 & 101 & HD 93222  & O7 V & O8 V \\
5     & 10:44:43.860 & -59:21:25.414 & 11.3  & 6.0 & 152 & HD 93249\tnote{a} & O9 III & \\
6     & 10:49:24.933 & -59:49:44.125 & 13.7  & 3.4& 132 & HD 305599 & B0 Ib & \\
\hline
\multicolumn{8}{c}{Other EROs} \\
7     & 10:38:34.415 & -59:43:10.861 & 11.6  &       &       &       &  & \\
8     & 10:40:19.370 & -59:49:09.479 & 13.8  &       &       &       &  & \\
9     & 10:40:29.175 & -58:53:23.162 & 7.6   & 7.2 & 144 &       &  & \\
10    & 10:40:49.119 & -59:50:27.150 & 11.2  &       &       &       &  & \\
11    & 10:40:50.810 & -58:52:27.934 & 11.6  &       &       &       &  & \\
12    & 10:41:06.765 & -58:50:06.036 & 7.0     &       &       &       &  & \\
13    & 10:42:45.146 & -59:52:19.672 & 12.1  &       & -12 & HD 305437 & B0.5 V & B0 V \\
14    & 10:43:09.463 & -59:24:53.87 & 11.8  &       &       &      & YSO\tnote{b} & \\
15    & 10:43:11.145 & -59:44:21.266 & 10.7  & 14.4 & 22 & HD 303316 & O6 V & O6 V \\
16    & 10:43:12.181 & -59:00:54.204 & 7.4   &       &       &       &  & \\
17    & 10:43:15.697 & -59:51:05.066 & 11.0    &       &       & HD 305516 & B0.5 V  & \\
18    & 10:43:20.289 & -60:13:01.348 & 59.3  & 15.5 & -91 & TYC 8957-99-1 & & B8 \\
19    & 10:43:43.771 & -58:42:20.749 & 16.8  & 1.6 & 36 &       &  & \\
20    & 10:43:43.959 & -59:48:18.132 & 12.8  & 15.4 & -102 & HD 305518 & O9.5 V & O9/O9.5 V \\
21    & 10:44:05.107 & -59:33:41.469 & 20.2  & 7.6 & 44 & Tr14-29 & B1.5 V\tnote{c} & B2 V \\
22    & 10:44:05.786 & -59:35:10.755 & 6.8   &       & 67 & Tr 14-124 & B1V & \\
23    & 10:44:30.184 & -59:26:12.723 & 7.4   & 6.9 & 144 & TYC 8626-2506-1 & O9 V(n)\tnote{d} & O9.5 V\tnote{b} \\
24    & 10:44:50.402 & -59:55:45.188 & 15.2  &       & -5 & CPD-59 2605 & B1 V & \\
25    & 10:45:13.359 & -59:57:54.092 & 15.3  &       &       & HD 305533  & B0.5 Vnn & B1 V \\
26    & 10:45:20.420 & -59:17:06.526 & 14.7  &       &       & HD 303300 & & O9.5 V\tnote{b} \\
27    & 10:45:38.315 & -60:36:15.023 & 10.8  &       &       &       & & \\
28    & 10:45:46.936 & -59:54:57.484 & 9.9   &       &       &       & YSO\tnote{b} & \\
29    & 10:46:11.525 & -58:39:11.410 & 16.8  & 5.2 & 14 & CPD-57 3781 & O8 V$^\text{e}$ & \\
30    & 10:46:13.585 & -59:58:31.381 & 10.7  &       &       &       &  & \\
31    & 10:46:53.817 & -60:04:42.114 & 19.8  & 4.7 & 68 & HD 93576  & O9 IV & O9 V \\
32    & 10:47:45.428 & -60:25:56.791 & 15.1  & 1.7 & -122 &       &  & \\
33    & 10:47:48.007 & -59:41:30.955 & 11.2  & 11.3 & 52 &       &  & \\
34    & 10:48:24.302 & -60:08:00.830 & 12.3  &       &       &       & YSO\tnote{b} & \\
35    & 10:51:11.555 & -59:48:44.916 & 18.2  &       &       &       &  & \\
\hline
\multicolumn{8}{c}{Extended 24~\um\ objects} \\
36    & 10:40:12.382 & -59:48:10.326 & 106.2 & 15.7 & -118 & HD 92607 & O8 V & O8.5 V+O9 V\tnote{f} \\
37    & 10:47:38.891 & -60:37:04.451 & 174.6 & 17.1 & 41 & HD 93683 & B0.5 Vne\tnote{g} & O9 V \\
38    & 10:47:46.173 & -60:24:36.173 & 25.4  & 7.3 & 25 &       & OB\tnote{b,h} & \\
39    & 10:48:46.555 & -60:35:40.461 & 27.6  & 5.8 & 16 & CPD-59 2735 & & B0 V \\
\bottomrule
\end{tabular}
\begin{tablenotes}
\footnotesize
\item[a]{HD 93249, or CD-58 3536, is the most luminous star in the Tr 15 cluster. It has a visual companion, CD-58 3536B O9.5 III \citep[see, e.g.,][and references therein]{CCCP_Tr15} whose wind likely also contributes to driving ERO 5.}
\item[b]{Candidate OB or YSO from \citet{Povich2,Povich3}.}
\item[c]{Identification and spectral type from \citet{Ascenso}. The driving star is identified as Trumpler 14 MJ 218 by \citet{Ngoumou}.}
\item[d]{New Galactic O Star Catalog spectral type from \citep{GOSS14}.}
\item[e]{Spectral type from \citet{Reed03}}
\item[f]{Double-lined spectroscopic binary.}
\item[g]{Be spectral type reported by \citet{Houk+Cowley}.}
\item[h]{Preliminary near-IR spectroscopy with the Southern Astrophysical Research (SOAR) telescope indicates an early-type star (M. Alexander, private communication, 2014).}
\end{tablenotes}
\end{minipage}
\end{threeparttable}
\end{table*}

\subsection{Size, Orientation, and Photometry}\label{sec:measurements}

We used the longest angular dimension of extended emission (be it
angular diameter for spheroidal objects or angular arc width for bow
shock structures) as a proxy to measure the angular size of each ERO
(Table~\ref{tab:EROdrivingstars}).  The location of the apparent
driving star was determined using the 3.6 $\mu$m images and the
Vela-Carina point source catalog.  Linearly scaled contours of the 8.0
$\mu$m (or 24~\um) intensity provided a means to determining shape of
the emission and locate the emission peak (Figure~\ref{fig:EROs}).
Drawing a straight line from the location of the star to the point of
highest 8.0 $\mu$m (or 24~\um) intensity provided measurements of both
the orientation angle and offset distance between the ERO and the
driving star (Table~\ref{tab:EROdrivingstars}).  For bow shock
candidates the offset distance gives the standoff distance between the
shock front and the star projected onto the plane of the sky (see
Figures~\ref{fig:EROs} and \ref{fig:24candidates}).

Twenty-one EROs (including all 24~\um\ objects) had measurable offset
distances and orientations, including 11 with no clear bow shock
morphology. These objects have both offset distances and orientations
listed in Table \ref{tab:EROdrivingstars}.
The contouring method also allowed us to measure orientations but not
offset distances for EROs 13, 22, and 24.  These three EROs are
characterized by 8.0 $\mu$m emission that is extended toward one side
of the star but not the other, but the contour peak overlaps with the
position of the star.  For the remaining 15 EROs, the highest
intensity of 8.0 $\mu$m emission fell within the PSF of the star;
these ``spherically symmetric'' EROs have no orientation listed in
Table \ref{tab:EROdrivingstars}.

We performed photometry on all 39 ERO candidates using IRAF.  To
ensure that the photometry apertures we drew were consistent in each
bandpass we convolved the IRAC images using a 2-D Gaussian with
$\sigma\approx2.6$ pixels to match the
$6''$ resolution of the MIPS images.  Median background subtraction
was performed on PSF-subtracted images using the IRAF DAOPHOT FITSKY
task \citep{Stetson}, which also provided a standard deviation in
counts for the background.
Using the DAOPHOT POLYMARK task, we constructed by hand irregularly
shaped apertures around each ERO using the convolved 8.0 $\mu$m images
(or the 24~\um\ images for EROs 36--39).  Background-subtracted flux
densities were then measured (in MJy/sr) using the IRAF POLYPHOT task.
We performed photometry on both the mosaic and residual images
produced by the GLIMPSE pipeline. Results from the mosaic photometry,
which include the flux contributions from the driving stars, are
reported in Table~\ref{tab:EROphotometry}.


\section{Results}


\subsection{Morphological Bow Shock Candidates}

P08 observed that bow shocks in both RCW~49 and M17 formed around late
O stars. S10 similarly found that EROs in Carina tend to form around
or near late O to early B-type stars.  EROs 1--6
(Figure~\ref{fig:EROs}) and 24~\um\ objects 36, 37 and 39 exhibit
arc-shaped morphology in the contour analysis consistent with bow
shocks (we refer to these as ``morphological bow shock candidates'' or
MBSc). All MBSc are associated with known late O or early B-type
driving stars (Table~\ref{tab:EROdrivingstars}).  EROs 1, 3, and 4
were previously identified by S10, but we do not classify the
remaining 5 S10 EROs as MBSc.
\begin{figure*}
\includegraphics[width=6.75in]{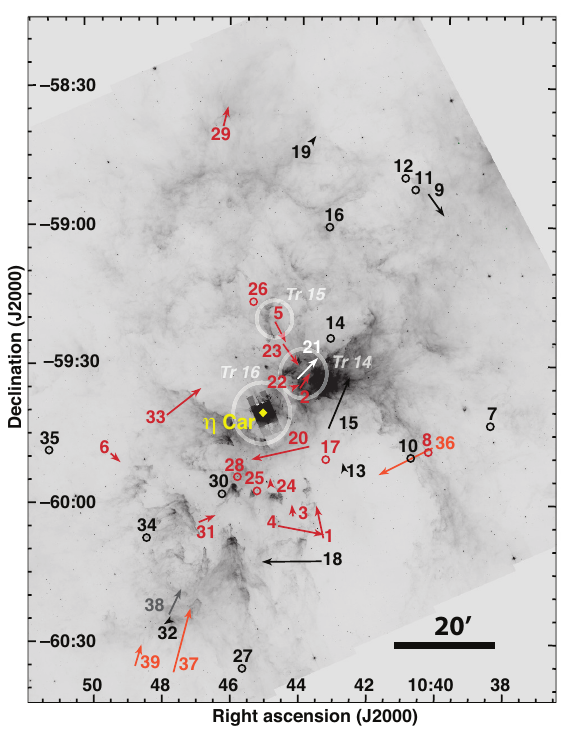}
\caption{Grayscale 8~\um\ image of the region of Carina Nebula
  searched for EROs. Morphological, 8~\um\ bow shock candidates (EROs
  1-6) are labelled as red arrows, other EROs (7-35) are black or
  white, 24.0 \um\ bow shocks (36, 37, and 39) are orange, while the
  other 24 \um\ objects (38) is gray. Lengths of the arrows correspond
  to approximate standoff distances scaled up by a factor of 50, and
  directions represent ERO orientations measured using intensity
  contours (Table \ref{tab:EROdrivingstars}).  Circles mark EROs
  without measurable orientations.  Locations of the principal
  ionizing clusters of the Carina Nebula, Tr 14, 15, and 16 are
  circled. 
}
\label{fig:CarinaMap}
\end{figure*}

Of the 9 MBSc, 6 show preferred ``inward'' orientations toward the
centre of the Carina Nebula (Figure \ref{fig:CarinaMap}).
EROS 1, 3, and 5 plus 24~\um\ objects 37 and 39 are all generally
oriented toward the central zone containing Tr 16 and Tr 14. ERO 2,
located in projection between Tr 16 and Tr 14, is oriented toward Tr
14.  EROs 4 and 6, plus 24~\um\ object 36, have ``transverse''
orientation, and none of the MBSc have the ``outward'' orientation
expected if the bow shocks were caused purely by the high space
velocity of a runway OB star escaping the Carina Nebula.


\subsection{Additional Bow Shock Candidates from Color Analysis of EROs}\label{sec:colors}

Because many EROs are barely resolved, IR colors offer the potential
for identifying additional bow shock candidates in the Carina Nebula
and other, more distant \hii regions.  Emission features from
polycyclic aromatic hydrocarbons (PAHs) tend to dominate broadband,
IRAC images of \hii\ regions \citep{Peeters}.  Most diffuse objects in
Carina have colors reflecting the strong PAH emission features at 3.3,
6.3, 7.7, and 8.6 $\mu$m, which fall within the 3.6, 5.8, and 8.0
$\mu$m IRAC filters, but there are no strong PAH features in the
4.5~\um\ filter \citep[][S10]{Povich2007}.

We produced color-color diagrams of EROs to investigate whether EROs
in general, and bow shock candidates in particular, exist in a color
space distinguishable from unresolved, nebular PAH knots and other
intrinsically red mid-IR point sources.

\begin{figure*}
\includegraphics[width=3.1in]{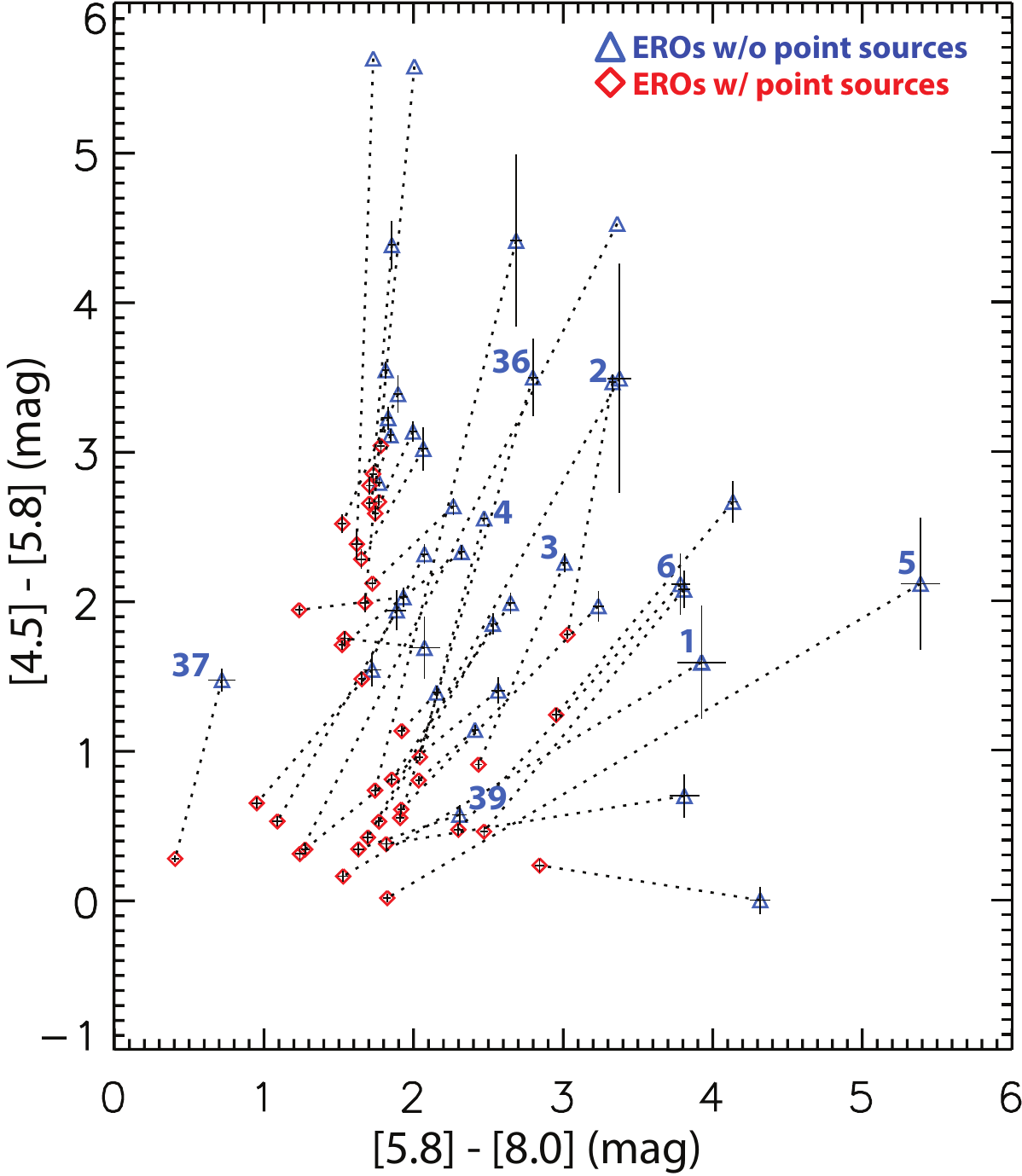}
\caption{[4.5]-[5.8] versus [5.8]-[8.0] color-color diagram of EROs in Carina. Each ERO is plotted twice, once with emission from the associated point source included in the color (diamonds) and once without point source emission (triangles) connected by dashed line segments. Morphological bow shock candidates are labelled by their ERO numbers. 
}
\label{fig:23vs34_EROs}
\end{figure*}

Figure \ref{fig:23vs34_EROs} shows the colors of EROs with and without
the contributions of their central stars to the IR flux.  All
photometry apertures were drawn to include the driving stars along
with the extended emission itself.  Not surprisingly, we find a
general trend for EROs (with point sources) to move redward in both
$[4.5]-[5.8]$ and $[5.8]-[8.0]$ when their point sources are
subtracted.
Uncertainties on colors significantly increase when point sources are
removed, mostly due to low signal-to-noise at 4.5 $\mu$m, as emission
in this bandpass is dominated by the candidate driving stars rather
than the EROs. For the remaining analysis, we employ ERO photometry
including the contributions of the associated driving stars
(Table~\ref{tab:EROphotometry}) because the stellar contribution will
be inevitably included for unresolved EROs.

\begin{figure*}
\includegraphics[scale=1]{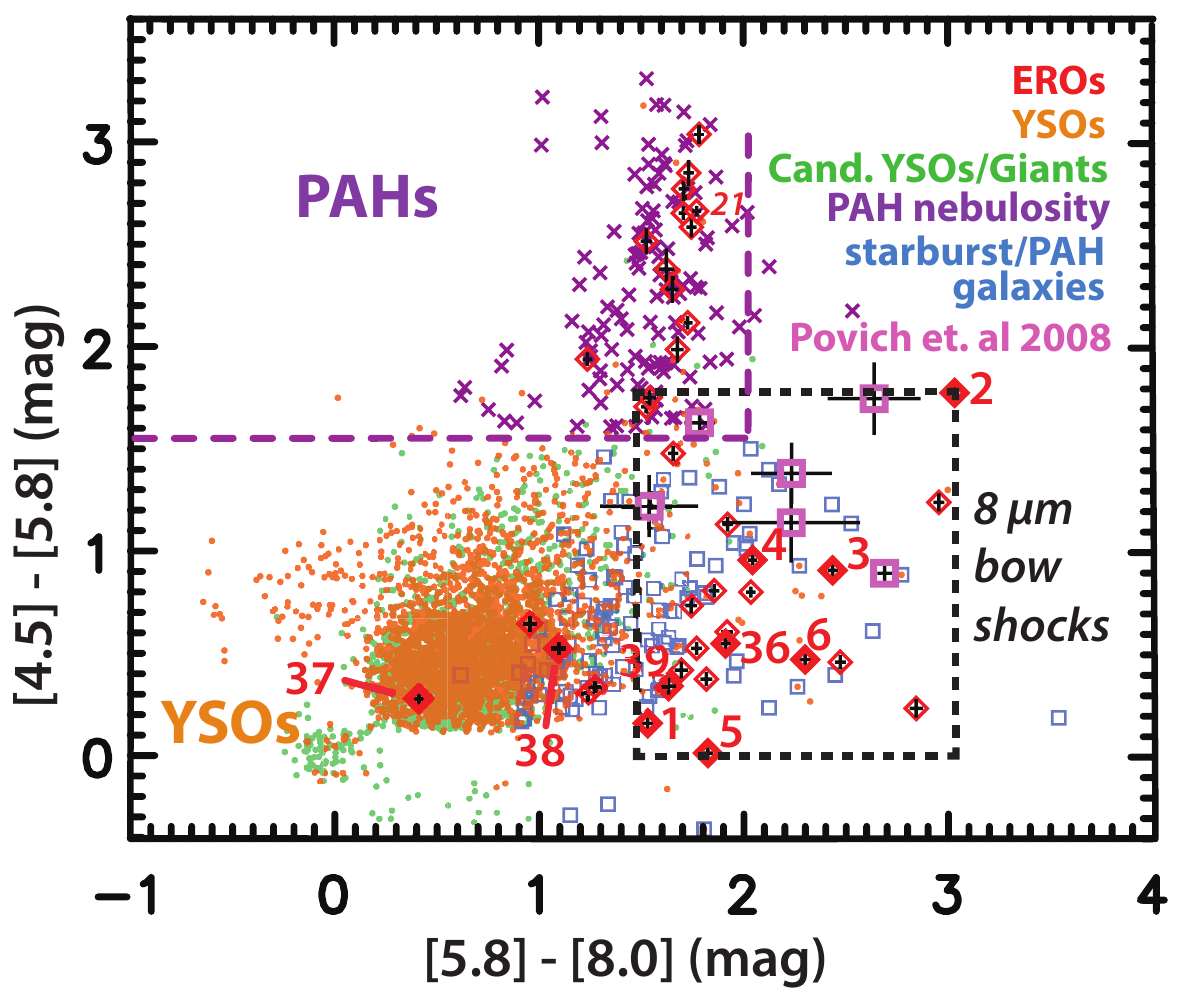}
\caption{[4.5]-[5.8] versus [5.8]-[8.0] color-color diagram of EROs presented in this paper, M17 and RCW49 bow shocks from P08, and various classes of intrinsically red IR point sources classified in the MIRES catalog \citep{Povich4}. Morphological and 24~\um\ bow shock candidates are labelled by their ERO number (Table~\ref{tab:EROdrivingstars}). Morphological 8~\um\ bow shocks are found within the region of color space bounded by the short-dashed black box. Colors are based on the photometry in Table~\ref{tab:EROphotometry} and include flux from the associated driving stars.}
\label{fig:23vs34_Povich_nonumbers}
\end{figure*}

We next compared the colors of EROs (with their central stars
included; Table~\ref{tab:EROphotometry}) to a subset of intrinsically
red IR point sources from the MYStIX IR Excess Source catalog
\citep[MIRES;][]{Povich4}. We defined our comparison sample as those
MIRES objects found in IRAC data processed through the GLIMPSE
pipeline and associated with massive star-forming regions similar to
the Carina Nebula, with distances ranging from 2--4~kpc. Using the
classifications of the MIRES catalog, we separate sources into three
primary regions on the $[4.5]-[5.8]$ vs.\ $[5.8]-[8.0]$ color plane
(Figure \ref{fig:23vs34_Povich_nonumbers}): (1) nebular PAH knots, (2)
young stellar objects (YSOs) with dusty, circumstellar
discs/envelopes, and (3) PAH-dominated emission from unresolved,
external starbursting galaxies. We find that 14 of the 39 EROs fall
within the PAH quadrant, while only 3 fall within the color space
designated for YSOs.  The remaining EROs reside within the color space
designated for starburst galaxies, however EROs are unlikely to be
confused with background galaxies because (1) these EROs are even
redder in
$[5.8]-[8.0]$ and (2) EROs associated with Galactic OB stars should be
much brighter in the mid-IR compared to unresolved, external galaxies.

EROs falling outside the PAH or YSO quadrants tend to occupy a
distinct, box-shaped region (dashed line in Figure
\ref{fig:23vs34_Povich_nonumbers}) defined by
\begin{displaymath}
1.5\le[5.8]-[8.0]\le 3.1~({\rm mag}),
\end{displaymath}
\begin{displaymath}
0\le[4.5]-[5.8]\le 1.8~({\rm mag}).
\end{displaymath}
This ``ERO box'' contains all of the 8~\um\ MBSc from both this paper
and P08 (12 total), plus two of the three 24~\um\ MBSc (ERO 37 is the
outlier).  It also contains 12 other EROs that we hereafter designate
``color bow shock candidates,'' or CBSc: EROs 8, 17, 20, 22, 23, 24,
25, 26, 28, 29, 31, and 33. Ten of these are associated with known OB
stars (Table~\ref{tab:EROdrivingstars}), including the remaining 5
EROS from S10. Of the 7 CBSc with measurable orientations, 5 are
oriented inward, similar to the MBSc.
\begin{figure*}
\includegraphics[scale=1]{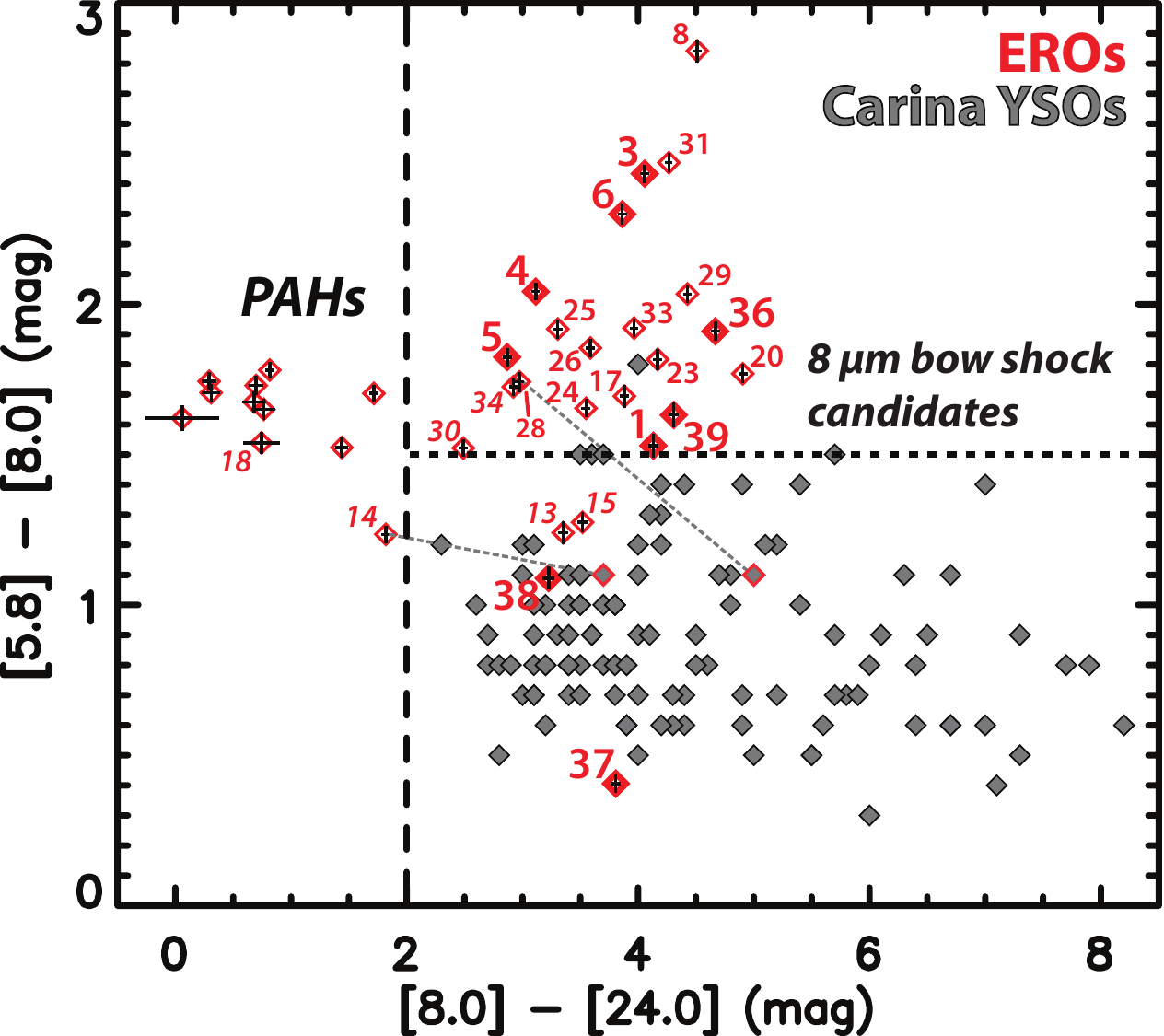}
\caption{$[5.8]-[8.0]$ versus $[8.0]-[24.0]$ color-color diagram comparing EROs to Carina YSOs from the PCYC \citep{Povich3}. Dashed lines separate regions of this color space containing YSOs, bow shock candidates, and PAH nebular knots. MBSc are labelled with bold numbers, CBSc are labelled with smaller numbers, and other notable EROs are labelled with italicized numbers. EROs 14 and 28 are associated with the point sources PCYC 179 and 990, respectively (joined by dashed gray lines). 
}
\label{fig:24vs45_YSOs}
\end{figure*}

In Figure \ref{fig:24vs45_YSOs} we compare the $[5.8]-[8.0]$ vs.\
$[8.0]-[24]$ colors of EROs (again including contributions from the
driving stars) to objects in the Pan-Carina YSO Catalog
\citep[PCYC;][]{Povich3}, which, unlike the MIRES catalog, includes
24.0~\um\ point-source photometry.
This color plane divides into 3 principal regions. Bow shock
candidates are again separated from most YSOs by the line at
$[5.8]-[8.0]=1.5$~mag defining one side of the ``bow shock box'' from
Figure~\ref{fig:23vs34_Povich_nonumbers}. EROs falling within the PAH
region of Figure~\ref{fig:23vs34_Povich_nonumbers} are much bluer than
bow shock candidates in $[8.0]-[24]$; providing another (loose)
constraint on CBSc colors:
\begin{displaymath}
  [8.0]-[24] > 2~({\rm mag}).
\end{displaymath}
All CBSc and MBSc are found in the range $2.7<[8.0]-[24]<5$~mag,
similar to disc-dominated YSOs in the PCYC. The PCYC also includes a
population of much redder sources (to $[8.0]-[24]\sim 8$~mag)
corresponding to embedded protostars \citep{Povich3}.

Four EROs fall within the YSO color region in both
Figures~\ref{fig:23vs34_Povich_nonumbers} and
\ref{fig:24vs45_YSOs}. ERO 37 is the 24~\um\ MBSc with anomalous
colors, perhaps the one true ``24~\um--only'' bow shock in our
sample. ERO 38 is the one 24~\um\ source with ambiguous morphology; it
may be a bow shock or part of a larger IR bubble structure around a
candidate OB star \citep{Povich2}.  ERO 13 is associated with a B0.5
star, and while its colors are consistent with a circumstellar disc,
its projected location in the middle of the Carina Nebula's southern
evacuated superbubble lobe (Figure~\ref{fig:CarinaMap}) seems to argue
against extreme youth, while its directly inward orientation favors a
bow shock interpretation. ERO 15 is associated with an O6 star, an
early type compared to other bow shock candidates; it may instead be
the brightest part of an IR bubble structure excited by that star.

Three EROs are the resolved 8~\um\ counterparts to PCYC point
sources. EROs 14 and 28 are associated with the point sources PCYC 179
and 990, respectively (Figure~\ref{fig:24vs45_YSOs}); both EROs are
redder in $[5.8]-[8.0]$ but much bluer (by ${\sim}2$~mag) in
$[8.0]-[24]$ compared to the point sources. This suggests that our
aperture photometry extracted preferentially more 8~\um\ flux compared
to the PCYC point-source photometry (which used the GLIMPSE pipeline
for the IRAC filters and a custom 24~\um\ aperture photometry
procedure; \citealp{Povich3}). Neither ERO 14 nor 28 has a discernible
orientation (Figure~\ref{fig:CarinaMap}); they are consistent with
YSOs with large, marginally resolved circumstellar discs.  ERO 34
corresponds to PCYC 1367, a {\it Midcourse Space Experiment} point
source (but not an IRAC point source, hence not among the PCYC sources
plotted in Figure~\ref{fig:24vs45_YSOs}). It is not associated with a
known OB star, but it may be a massive YSO \citep{Povich3}.

\section{Discussion}

\subsection{Bow Shocks as ISM ``Weather Vanes''}
Bow shock candidates (MBSc and CBSc) associated with OB stars are of
particular interest as probes of massive star feedback processes and
the relative motions between massive stars and the ISM in \hii
regions.  EROs 1, 3, 24, 31, 37 and 39, all located in the South
Pillars region, are generally aligned with the various pillars (S10)
and oriented in the inward direction toward Tr 14 \& 16
(Figure~\ref{fig:CarinaMap}). This preferred orientation suggests that
these bow shocks are caused by the interaction of the driving OB
stellar wind with with the global feedback processes (thermal gas
pressure, radiation pressure, and/or cluster winds) powering the
expansion of the \hii region and eroding the pillars.  EROs 2, 22, and
23 are all in closer proximity to Tr 14 than to Tr 16, and are all
oriented toward the centre of the Tr 14 cluster, indicating that they
are reacting to feedback from Tr 14 only. No EROs appear oriented
toward Tr 16 but not Tr 14; this may be due to the fact that Tr 14 is
a very dense, centrally concentrated cluster while Tr 16 is a looser
aggregate of multiple subclusters \citep{CCCP_clustering}.

ERO 5 is a particularly interesting case. It is apparently driven by
the visual binary stars HD 93249 (or CD--58 3536A) and CD--58 3536B,
the most luminous members of the Tr 15 cluster \citep{CCCP_Tr15}. ERO
5 is oriented directly toward the centre of the Tr 14 cluster,
however, providing direct evidence that feedback from Tr 14 completely
overwhelms that of the significantly older, less massive and less
luminous Tr 15. This evidence of physical interaction between the
clusters provides further support that both clusters are at the same
heliocentric distance and hence part of the Carina Nebula Complex.

Five bow shock candidates assoicated with OB stars, EROs 4, 6, 20, 29,
and 36, are not oriented toward the centre of the Carina Nebula. These
divergent orientations have two possible causes: (1) locally
non-radial components to the ambient ISM flow, perhaps produced by
photoevaporative flows off of nearby molecular clouds (S10), or (2)
high space velocity of the driving star, as with the candidate
``runaway'' OB stars identified via IR bow shocks found near other
massive Galactic star-forming regions
\citep{RunawaysI,Kobulnicky,RunawaysII}.
ERO 4 is perhaps the best candidate for a bow shock interacting with a photoevaporative flow, as
it is oriented toward the inner side of the large pillar demarcating
the western boundary of the South Pillars region. ERO 29 is a
candidate runaway O8 V star, as it is located on the northern
periphery of the Carina Nebula and is oriented almost precisely
outward from the centre (Figure~\ref{fig:CarinaMap}).  
To confirm the runaway star scenario requires a better understanding of the complex kinematics of the Carina stellar population. Interpreting velocity measurements for individual OB stars begins with defining a reliable reference frame, correcting for Galactic rotation and streaming motions, as well as accounting for cluster dynamics.  Additional radial velocity and proper motion data should give further insight to the potential runaway status of the Carina bow shock driving stars in the future.



\subsection{Physical Conditions Governing Emission Processes and
  Colors of IR Bow Shocks}

Our analysis of the IR colors of EROs (Section~\ref{sec:colors})
supports the conclusions of P08, S10, and \citet{Kobulnicky} that IR
emission from bow shocks associated with OB stars in \hii regions
originates from dust entrained within the (pre-shock) plasma and
heated by the nearby OB star (P08).  Because O and B-type stars are
are not efficient dust producers, we echo S10 in proposing that the
source of dust is (photo)evaporating molecular clouds, such as the
molecular cloud formations being ablated in the South Pillars
region. \citet{Everett} calculated that dust on the order of
$a\le0.1$~\um\ will typically be evacuated from a wind-blown \hii
region cavity on time-scales of $10^4$ years, and any dust grains on
the order of $a\simeq0.001$~\um\ will be destroyed, hence dust must be
continually replenished and likely does not survive passage through
the stellar-wind bow shocks.  The wind momenta of massive stars places
additional constraints on the types of stars that can drive IR bow
shocks, because the winds cannot be so strong such that they
completely clear away the surrounding dust and gas (P08, S10). This
interpretation favors IR bow shocks forming around late O and early B
stars on the periphery of young, dusty, star-forming regions, fitting
the general picture of the Carina Nebula (Figure~\ref{fig:CarinaMap}).

We have demonstrated that dusty bow shocks associated with early-type
stars generally occupy regions of IR color space that are distinct
from the colors of the most likely contaminating sources, PAH nebular
knots and YSOs. PAH knots are redder in $[4.5]-[8.0]$ but bluer in
$[8.0]-[24]$ compared to CBSc. This is explained by the presence of
strong PAH emission features falling within the [5.8] and [8.0]
bandpasses but not in the [4.5] or [24] bandpasses. Evidently IR bow
shocks associated with OB stars do not excite strong PAH emission, and
this explains why EROs are visually distinguishable from the ambient
PAH nebulosity in our multiband \spitzer\ color images
(Figure~\ref{fig:EROs}). There are several possible physical reasons
for this dearth of PAH emission: (1) the dust entrained within the bow
shocks was PAH-depleted already, typical for dust within \hii regions
\citep{Povich2007,Everett}, (2) PAH molecules mixed within the
pre-shock gas and dust are destroyed by FUV radiation from the OB
driving star, or (3) PAH molecules are destroyed by the passage
through the shock front itself.

CBSc are redder than YSOs in $[5.8]-[8.0]$
(Figure~\ref{fig:23vs34_Povich_nonumbers}), but the $[8.0]-[24]$
colors of CBSc are similar to disc-dominated YSOs, and protostars
extend to much redder $[8.0]-[24]$ colors
(Figure~\ref{fig:24vs45_YSOs}). This indicates that the emitting dust
in bow shocks is generally warmer than the dust in protoplanetary
discs and protostellar envelopes. Another difference is that YSOs,
especially protostars, often exhibit deep silicate absorption at
$9.7$~\um, which suppresses flux in the [8.0] band. This absorption is
not expected to be present in IR bow shocks, which should be optically
thin (P08) and possibly exhibit the silicate feature in {\em
  emission}.

\subsection{Probing ISM Pressure and Velocity in the Carina Nebula}

Following P08, we quantify the physical parameters governing the
observed bow shock morphology to measure the momentum flux of the ambient
ISM at different locations within the Carina Nebula.
We calculate the ambient momentum flux of the ambient ISM at the standoff distance $d_w$ of the bow shock, that is, the point at which momentum flux is balanced by the driving star's stellar wind, as
\begin{displaymath}
n_wv_w^2=n_0v_0^2,
\end{displaymath}
where $v_0$ is the {\em relative} velocity between the driving star
and the ambient ISM.  We normalize the values for mean ISM particle
density, stellar mass-loss rates, and terminal wind velocities such
that $n_{0,3}=n_0/(10^{3}$ cm$^{-3}$),
$\dot{M}_{w,-6}=\dot{M}_w/(10^{-6}$ M$_\odot$ yr$^{-1}$), and
$v_{w,8}=v_w/(10^8$ cm s$^{-1}$), following \citet{vanBuren2}, P08,
and \citet{Kobulnicky}. Assuming that mass-loss about the star is
spherically symmetric,
it can be written as
\begin{displaymath}
\dot{M}_w=4\pi d_w^2 \mu n_w v_w,
\end{displaymath}
where $\mu=2.36\times10^{-24}$ g is the mean ISM gas mass per hydrogen
atom.  Solving for $v_0$, we have
\begin{equation}
v_0=1.5\left(\frac{d_w}{\text{pc}}\right)^{-1}\left(\dot{M}_{w,-6}v_{w,8}\right)^{1/2}n_{0,3}^{-1/2}\quad\left[\text{km s}^{-1}\right].
\end{equation}
The resulting values for $v_0n_{0,3}^{1/2}$ derived from each of the
MBSc, are presented in Table~\ref{tab:BowShockDrivingStars}. These are
comparable to values cited by P08, indicating that overall the
MBSc in the Carina Nebula probe similar ISM conditions as found in M17
and RCW 49. As found by \citet{Fullerton}, and noted in P08, mass-loss
rates have a high dependence on spectral type and are therefore the
largest source of uncertainty in our calculations. Indeed, citing the
high uncertainty in mass-loss rates, \citet{Kobulnicky} turned the
problem around, assuming reasonable values for the ambient density and
relative velocities of OB stars to the ISM to use bow shocks as
constraints on the stellar wind momentum and hence mass-loss rates.

Equation 1 must be interpreted with caution. Order-of-magnitude
uncertainties in mass-loss rates \citep{Kobulnicky} create factors of
${\sim}2$ uncertainty in $v_0$. All results are modulo an unknown
inclination angle with respect to the plane of the sky (although
$\cos{i}$ is likely to be near unity for MBSc, otherwise the arc shape
would not be visually apparent, see P08). The density of ionized
plasma in the Carina Nebula likely varies by factors of several about
the typical value of $n_{0,3}=1$ \citep{Brooks}, but this effect is
subordinate to the uncertainties in the mass-loss rates. More
importantly, the derived values for $v_0$ (assuming $n_{0,3}=1$) are
comparable to the ${\sim}10$~km~$^{-1}$ sound speed in the ionized
plasma. In the case where the relative velocity between the star and
the ambient ISM is not highly supersonic, thermal (and possibly
turbulent) pressure in the ambient medium become important relative to
ram pressure in balancing the stellar wind pressure. Accounting for
thermal pressure adds a term, proportional to the square of the sound
speed, to our momentum flux balance equation, which becomes
\begin{displaymath}
  n_wv_w^2 = n_0v_0^2 + 2\frac{n_0}{\mu}k_B T_0
\end{displaymath}
(note that the wind itself is highly supersonic, hence the thermal
pressure term on the left-hand side can still be safely
neglected). Normalizing the ambient temperature to
$T_{0,4}=T_0/(10^4$~K), we derive a correction to the relative
velocity from Equation 1,
\begin{equation}
  v'_0 = \left(v_0^2 - 117~T_{0,4}\right)^{1/2}.
\end{equation}
This correction is independent of the ambient density, and for \hii
regions $T_{0,4}$ is near unity. We can thus apply Equation 2 to
correct our values of $v_0n_{0,3}$ for the expected contribution of
thermal pressure, except for two cases (EROs 1 and 37) where the
thermal pressure term exceeds the ram pressure term
($v_0<\sqrt{117}$~km~s$^{-1}$). Corrected values of $v'_0n_{0,3}$ are
included in Table~\ref{tab:BowShockDrivingStars}; in the majority of
cases these corrections are small compared to the uncertainties in
$v_0$.

S10 used the observed bow shock standoff distances to contrain the
{\it total} pressure in the ambient medium balanced by the stellar
wind pressure,
\begin{equation}
  p_0 = \frac{1}{13}\left(\frac{d_w}{\text{pc}}\right)^{-2}\dot{M}_{w,-6}v_{w,8}\quad\left[10^{-9}~\text{dyne cm}^{-2}\right].
\end{equation}
This approach does not assume any particular geometry for the shock
front, does not require a distinction of ram pressure from other
sources of ambient pressure, and does not depend on the density or
temperature of the ambient gas. Equation 3 is, however, more sensitive
to uncertainties in mass-loss rates and standoff distances than
Equation 1. We list values for $p_{0,-9}=p_0/(10^{-9}$ dyne cm$^{-2}$)
in Table~\ref{tab:BowShockDrivingStars}, these are consistent with the
values of $p_{0,-9}=1.8$, 4.8, and 8.6 derived for EROs 1, 3 and 4,
respectively, by S10, the differences reflecting the adoption of
different mass-loss rates for the driving stars.




\begin{figure*}[th]
\centering
	\begin{subfigure}[b]{0.4\textwidth}
	\includegraphics[scale=0.65]{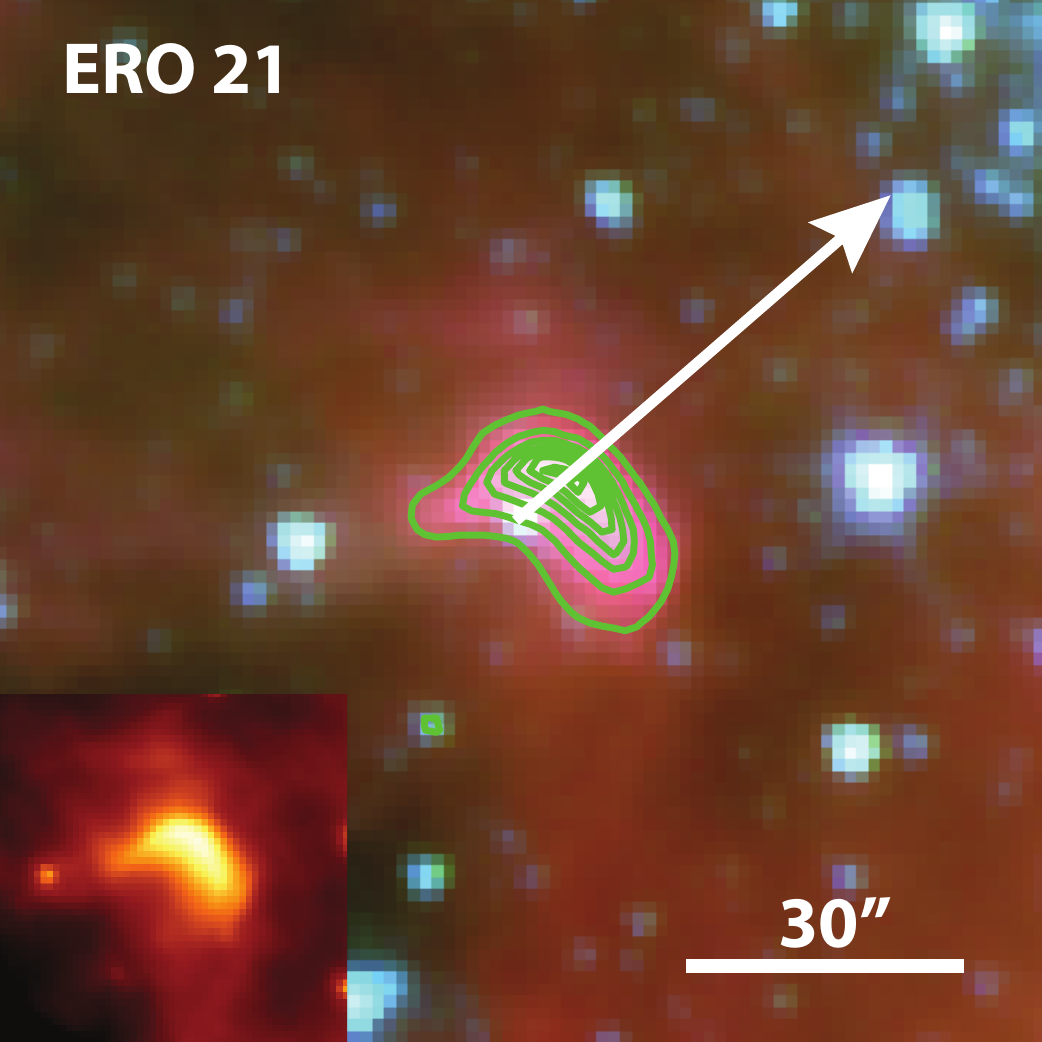}
	\end{subfigure}
	\begin{subfigure}[b]{0.4\textwidth}
	\includegraphics[scale=0.65]{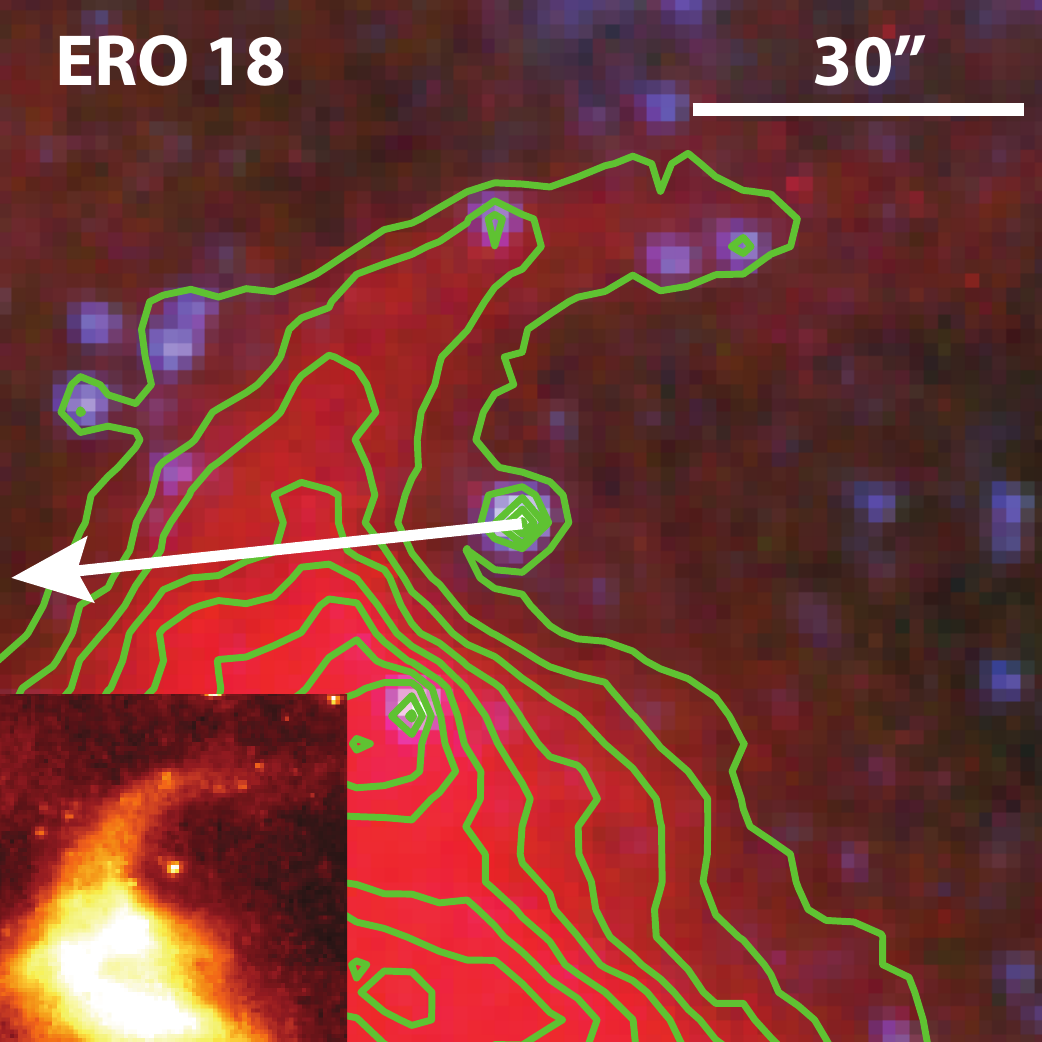}
	\end{subfigure}
\caption{Composite 3-color images of ERO 21 (compact nebula from \citealp{Ascenso}) and ERO 18 with the same color-coding, inset images, contours, and overlays as in Figure~\ref{fig:EROs}.
}
\label{fig:other_EROs}
\end{figure*}


\begin{table*}
\begin{minipage}{130mm}
\caption{\label{tab:BowShockDrivingStars} Bow shock standoff distances and estimated stellar wind properties.  Estimates of stellar mass-loss rates and
  wind-velocities are based on \citet{Vink}, \citet{Brooks},
  \citet{Fullerton}, \citet{Smith2006b}, \citet{Muijres2012a}, and
  \citet{Muijres2012b}.}
\begin{tabular}{rcccccc}
\toprule[1.5pt]
ERO & Spectral Type & $\dot{M}_{w,-6}v_{w,8}$ & $d_w\cos i$ & $v_0n_{0,3}^{1/2}(\cos i)^{-1}$ & $v'_0n_{0,3}^{1/2}(\cos i)^{-1}$ & $p_{0,-9}$\\
& & & (pc) & (km s$^{-1}$) & (km s$^{-1}$) & \\
\midrule
1     & O9.5 IV & $\simeq$0.2   & 0.096  & 7 & \nodata & 1.7 \\
2     & O9 V  & $\simeq$0.2   & 0.053 & 13 & 7 & 5.5 \\ 
3     & O9.5 V & $\simeq$0.2   & 0.034 & 20 & 17 & 13 \\
4     & O7 V  & 0.5--2.5 & 0.13 & 8--18 & ${\le}14$ & 2.3--11 \\
5     & O9 III+O9.5 III    & 1--3  & 0.067 & 22--39 & 19--37 & 17--51 \\ 
6     & B0 Ib & 0.5--2.5 & 0.038  & 30--62 & 28--61 & 27--133 \\
\hline
36    & O8.5 V+O9 V  & 1--5 & 0.17 & 9--20 & ${\le}17$ & 2.7--13 \\
37    & O9 V  & $\simeq$0.2   & 0.19 & 3.5 & \nodata & 0.43 \\
39    & B0 V  & $\simeq$0.2   & 0.064 & 11 & 2 & 3.8 \\
\bottomrule
\end{tabular}
\end{minipage}
\end{table*}

\subsection{Anomalous, Arc-Shaped EROs}
Two notable EROs have arc-shaped morphology but are not counted among
our MBSc.  

ERO 21, identified by \citet{Ascenso} as a compact nebula
with bow shock morphology in VLT $JHK_s$ images (and included in the
S10 list of EROs as S9, although it fell outside the boundary of the
IRAC images analysed in that work), also appears arc-shaped in the
Vela--Carina IRAC images (Figure \ref{fig:other_EROs}).  Contour
analysis of ERO 21 confirms that peak emission at 8.0 $\mu$m occurs
between the direct line of sight between its driving star and the
centre of Tr 14. 
The $[4.5]-[5.8]$ and $[5.8]-[8.0]$ colors of ERO 21 place it deep
within the region dominated by PAH nebulosity (Figure
\ref{fig:23vs34_Povich_nonumbers}), and these anomalous colors prevent
us from classifying this object as an MBSc.  The original \citet{Ascenso}
classification of ERO 21 as a ``compact nebula" rather than a dusty
bow shock appears appropriate, as its mid-IR colors indicate that this
object is likely the tip of an elongated column of high-density gas
and dust, similar to the numerous star-forming pillars observed
elsewhere in the Carina Nebula. ERO 21 is, however, driven by an early
B1.5 V star, which are known to have strong winds associated with
stellar wind bow shocks, so it is certainly possible that an
irradiated shock front has formed at the tip of the pillar.  We
measured a standoff distance of 0.084 pc, similar to the 8.0~\um\ MBSc
listed in Table \ref{tab:BowShockDrivingStars}, giving ambient ISM
properties around ERO 21 of $v_0n_{0,3}^{1/2}\approx 4$ and
$p_{0,-9}=2.2$, both on the low range compared to other EROs,
especially the nearby (in projection) ERO 2
(Figure~\ref{fig:CarinaMap}).

\citet{Ngoumou} named ERO 21 ``the sickle'' and proposed a model in which the wind of a
runaway B star interacts with an ambient density gradient in a low-density envelope surrounding a nearby, compact clump of molecular material. 
The bow-shock hence forms  as a consequence of the high velocity of the driving star, Trumpler 14 MJ 218, which has a UCAC proper motion of $95$~\kms\ at position angle $37^{\circ}$ (W of N).
 If this interpretation is correct, the unusual, PAH-dominated colors of ERO 21 might be explained by the
proximity of MJ 218 to a molecular clump of a nearby pillar, which is externally illuminated by either the B star itself or the nearby O stars in Tr 14. Unlike our other bow shock candidates, the globule must be sufficiently dense either to shield PAH molecules from destruction or to provide a sufficient supply of PAH-rich gas and dust to replenish IR-emitting arc. While the uncertainties are quite large ${\sim}50\%$, the proper motion is consistent with the orientation of ERO 21 (Table~\ref{tab:EROdrivingstars}), which is the strongest evidence favoring the runaway star bow shock interpretation.

The runway star interpretation, however, has some important caveats. 
\citet{Ngoumou} note that the proper motion of MJ 218 is consistent with a runaway B star from Tr 16. They also note that the star is detected in X-rays \citep{CCCP_Cat}, and explain (reasonably) that the X-rays are produced by a low-mass, binary companion to the B star. It is not clear that such a binary system could survive a dynamical interaction (such as an N-body encounter or supernova explosion of the most massive star in an originally triple or higher-multiplet system) capable of ejecting it at ${\sim}100$~\kms\ from Tr 16. It is possible that the proper motion measurements are unreliable, given that the large distance to Carina, crowded field, and high background nebulosity conspire to make automated proper motion measurements very difficult.  The choice of a reliable zero-velocity reference frame is again very important.  The implicit assumption that the Carina Nebula as a whole has zero proper motion may apply thanks to its proximity to a Galactic rotation tangent point, but it is far from clear that the assumption of zero proper motion for Tr 16 and other individual (sub)clusters is valid.  The {\em Chandra} Carina Complex Project examined UCAC3 and other proper motion catalogs and concluded that the data as a whole were of insufficient quality to aid in the interpretation of the X-ray source population \citep{CCCP_Cat}. Finally, we note that a high proper motion star is not necessary to explain the morphology of ERO 21, as feedback-driven flows from the nearby Tr 14 cluster could also sculpt the associated globule into the observed ``sickle'' shape.

ERO 18 was previously identified by S10 (their ERO S3).  The
morphology and mid-IR colors of ERO 18 suggest that it may be a wind-carved
cavity in a larger molecular cloud rather than a bow shock (Figure
\ref{fig:other_EROs}). ERO 18 presents a special case for the
orientation and offset distance measurements
(Section~\ref{sec:measurements}); because the arc is contiguous with a
larger nebular structure we bisected the visible arc in lieu of using
the position of peak 8~\um\ emission.
As with ERO 21, the colors of ERO 18 are consistent with PAH-dominated
emission (Figure~\ref{fig:24vs45_YSOs}).  Our 2dF/AAOmega spectra give
a B8 type for the star apparently associated with ERO 18, but this may
be a chance alignment.  A late B-star is unlikely to provide
sufficient wind momentum to form a detectable ionized shock front,
particularly one with such a large standoff distance as ERO 18.  The
driving star also has proper motion reported in the UCAC4 catalog equivalent to ${\sim}30$~\kms\  at position angle $65^{\circ}$ W of N
\citep{UCAC4}. This proper motion is not significant given the reported errors, and in any case points in the {\em opposite} direction of the orientation of ERO 18 (Table~\ref{tab:EROdrivingstars} and Figure \ref{fig:other_EROs}). For these reasons we do not
classify ERO 18 and the apparently associated late B star as an MBSc.

\section{Summary}

We have identified 39 EROs in the Carina Nebula through visual
inspection of \spitzer\ mid-IR images. Among these EROs are 17
candidate IR bow shocks associated with OB stars, 6 appearing as
8.0~\um\ arcs, 3 as 24~\um\ arcs, and the remaining 8 identified using
their mid-IR colors.  Six of these bow shock candidates were
previously identified by S10, while the IR colors and morphologies of
the remaining 3 EROs cataloged by S10 (their S1, S3, and S9---the
``compact nebula" reported by \citealp{Ascenso}) favor PAH-dominated
nebular structures over bow shocks.

The majority of bow shock candidates (10) are oriented inward toward
the central ionizing clusters of the Carina Nebula
(Figure~\ref{fig:CarinaMap}). These are examples of {\em in situ} bow
shocks where the winds of late O/early B stars interact with ambient
ISM flows driven by the global expansion of the \hii region, as
observed in M17 and RCW 49 by P08. Five bow shock candidates have
transverse or outward orientations that could indicate high space
velocities consistent with runaway OB stars, as previously observed in
NGC 6611 \citep{RunawaysI}, Cygnus-X \citep{Kobulnicky}, and NGC 6357
\citep{RunawaysII}, or locally non-radial flows of plasma
photoevaporating off of nearby molecular cloud surfaces.  Analysis of
measured standoff distances demonstrates that the ambient momentum
flux in the expanding Carina \hii region is comparable to the M17 and
RCW 49 \hii regions (P08).

We find that all 8.0~\um\ bow shock candidates occupy a box-shaped
region on the $[4.5]-[5.8]$ vs.\ $[5.8]-[8.0]$ color plane that is
largely free of nebular PAH knots, YSOs, and PAH-dominated external
galaxies (Figure~\ref{fig:23vs34_Povich_nonumbers}). We further find
that all candidate bow shocks have $[8.0]-[24] > 2$~mag, providing an
additional means of separating them from PAH knots
(Figure~\ref{fig:24vs45_YSOs}).  Because the mid-IR extinction law is
observed to be approximately flat from the [4.5] through [24]
\spitzer\ filters \citep{Flaherty}, these color cuts are conveniently
reddening-free.  These cuts provide a simple prescription for
identifying candidate unresolved bow shocks based on MIR colors. It is
possible, for example, that some of the reddest sources classified as
YSOs in the MIRES catalog \citep{Povich4} are actually unresolved bow
shocks.

\section*{Acknowledgements} 
We thank the anonymous referee for their positive and useful suggestions to help improve this work.
We thank S. R. Majewski and R. Indebetouw for
providing early access to advanced data products from the Vela--Carina
survey and M. V. McSwain for reducing the 2dF/AAOmega spectroscopic
data and performing the spectral classifications for OB stars
associated with EROs. We gratefully acknowledge support from NSF award
AST-0847170 (CAMPARE; ALR, PI). NS received partial support from NASA
through awards issued by JPL/Caltech as part of GO programs 3420,
20452, and 30848. This work is based on observations from the {\em
  Spitzer Space Telescope} GO programs 30848 (MIPSCAR; N. Smith, PI)
and 40791 (Vela–Carina; S. R. Majewski, PI). {\em Spitzer} is operated
by the Jet Propulsion Laboratory, California Institute of Technology
through a contract from NASA.



\onecolumn
\begin{landscape}
\begin{table*}
\caption{\label{tab:EROphotometry} Total IRAC and MIPS mid-IR flux densities for EROs and probable driving stars (Jy).  Reported IRAC and MIPS flux densities $F$ are background-subtracted.  Irregularly shaped apertures were used to measure fluxes were drawn by hand to outline the ERO emission visible in the 8.0 $\mu$m residual images.  The 8.0 $\mu$m source apertures were used to measure fluxes at all other bandpasses. Separate circular apertures were used to determine the median background, then the background flux density $B$ was computed for the corresponding source aperture.  EROs 36-39 are 24.0 $\mu$m objects, for which source apertures were drawn using the 24.0 $\mu$m emission.  EROs 2, 21, and 22 do not have 24.0 $\mu$m data because they fall in saturated regions of the MIPS mosaic.}
\begin{tabular}{rrrcrrcrrcrrcrrrc}
\toprule[1.5pt]
ERO &  \multicolumn{12}{c}{IRAC} & & \multicolumn{3}{c}{MIPS} \\
  \cline{2-13} \cline{15-17}
& $F_{[3.6]}$ & $\delta_{[3.6]}$ & $F_{[3.6]}/B_{[3.6]}$ &  $F_{[4.5]}$ & $\delta_{[4.5]}$ & $F_{[4.5]}/B_{[4.5]}$ &  $F_{[5.8]}$ & $\delta_{[5.8]}$ & $F_{[5.8]}/B_{[5.8]}$ & $F_{[8.0]}$ & $\delta_{[8.0]}$ & $F_{[8.0]}/B_{[8.0]}$ & &  $F_{[24.0]}$ & $\delta_{[24.0]}$ & $F_{[24.0]}/B_{[24.0]}$ \\
\midrule
1     & 11.4  & 0.2   & 2310  & 7.2   & 0.2   & 1804  & 5.3   & 0.1   & 412   & 12.3  & 0.2   & 2486  && 61.1  & 0.4   & 190 \\
2     & 9.9   & 0.2   & 372   & 9.9   & 0.2   & 303   & 32.4  & 0.1   & 219   & 298   & 1     & 11195 &&       &       &  \\
3     & 12.7  & 0.3   & 1287  & 9.6   & 0.2   & 796   & 14.2  & 0.1   & 562   & 75.1  & 0.5   & 7598  && 348   & 1     & 365 \\
4     & 84.4  & 0.9   & 3187  & 63.6  & 0.7   & 2577  & 98.2  & 0.4   & 1125  & 363   & 1     & 13726 && 707   & 2     & 937 \\
5     & 16.3  & 0.2   & 1589  & 10.6  & 0.1   & 1264  & 6.9   & 0.1   & 144   & 20.8  & 0.2   & 2033  && 32.3  & 0.2   & 55 \\
6     & 16.6  & 0.4   & 1678  & 11.5  & 0.3   & 1361  & 11.3  & 0.2   & 341   & 53.0  & 0.6   & 5359  && 205   & 1     & 2173 \\
\hline
7     & 1.2   & 0.1   & 280   & 1.0   & 0.1   & 236   & 5.14  & 0.06  & 422   & 13.3  & 0.2   & 3159  && 3.0   & 0.2   & 37 \\
8     & 2.2   & 0.1   & 533   & 1.9   & 0.1   & 475   & 1.47  & 0.03  & 87    & 11.4  & 0.1   & 2715  && 80.3  & 0.2   & 748 \\
9     & 1.9   & 0.1   & 372   & 2.08  & 0.05  & 549   & 2.41  & 0.03  & 78    & 3.3   & 0.1   & 626   && 0.2   & 0.1   & 11 \\
10    & 1.4   & 0.1   & 216   & 1.11  & 0.04  & 180   & 7.7   & 0.1   & 236   & 21.5  & 0.2   & 3214  && 3.1   & 0.1   & 32 \\
11    & 0.53  & 0.03  & 105   & 0.55  & 0.03  & 157   & 3.57  & 0.04  & 135   & 8.2   & 0.1   & 1616  && 3.4   & 0.1   & 137 \\
12    & 0.62  & 0.05  & 198   & 0.39  & 0.03  & 157   & 2.25  & 0.04  & 174   & 5.7   & 0.1   & 1808  && 0.7   & 0.2   & 30 \\
13    & 4.9   & 0.1   & 1839  & 3.30  & 0.07  & 1561  & 2.80  & 0.04  & 270   & 5.0   & 0.1   & 1854  && 12.0  & 0.1   & 194 \\
14    & 1.5   & 0.1   & 213   & 4.8   & 0.1   & 692   & 18.2  & 0.1   & 527   & 32.0  & 0.2   & 4487  && 18.9  & 0.2   & 72 \\
15    & 22.0  & 0.4   & 1923  & 14.8  & 0.3   & 1185  & 12.9  & 0.1   & 325   & 23.5  & 0.3   & 2053  && 66.5  & 0.5   & 122 \\
16    & 0.57  & 0.03  & 94    & 0.53  & 0.03  & 132   & 2.10  & 0.03  & 61    & 5.6   & 0.1   & 916   && 1.1   & 0.1   & 36 \\
17    & 3.6   & 0.1   & 935   & 2.60  & 0.07  & 644   & 2.45  & 0.04  & 218   & 6.6   & 0.1   & 1704  && 25.9  & 0.2   & 132 \\
18    & 16.3  & 0.5   & 1196  & 9.4   & 0.4   & 925   & 30.3  & 0.4   & 742   & 70.5  & 1.1   & 5162  && 15    & 2     & 39 \\
19    & 1.7   & 0.1   & 258   & 1.3   & 0.1   & 265   & 10.7  & 0.1   & 288   & 29.0  & 0.2   & 4301  && 4.3   & 0.2   & 127 \\
20    & 16.4  & 0.3   & 1331  & 11.2  & 0.2   & 799   & 11.6  & 0.1   & 266   & 33.4  & 0.3   & 2707  && 338   & 1     & 418 \\
21    & 26.9  & 0.3   & 672   & 19.8  & 0.2   & 476   & 146.8 & 0.3   & 705   & 422   & 1     & 10556 &&       &       &  \\
22    & 0.92  & 0.03  & 104   & 0.79  & 0.03  & 83    & 1.58  & 0.02  & 33    & 13.5  & 0.1   & 1534  &&       &       &  \\
23    & 17.3  & 0.4   & 820   & 12.3  & 0.3   & 619   & 11.1  & 0.1   & 109   & 33.5  & 0.4   & 1587  && 172   & 1     & 250 \\
24    & 3.0   & 0.1   & 295   & 3.2   & 0.1   & 263   & 8.0   & 0.1   & 292   & 20.6  & 0.2   & 2050  && 59.9  & 0.2   & 239 \\
25    & 4.4   & 0.1   & 524   & 3.50  & 0.08  & 378   & 3.91  & 0.04  & 133   & 12.9  & 0.1   & 1546  && 29.9  & 0.2   & 108 \\
26    & 7.6   & 0.2   & 1308  & 6.5   & 0.1   & 1200  & 8.7   & 0.1   & 317   & 27.1  & 0.2   & 4696  && 81.6  & 0.3   & 801 \\
27    & 1.0   & 0.1   & 144   & 0.80  & 0.04  & 159   & 7.1   & 0.1   & 235   & 19.7  & 0.2   & 2968  && 4.1   & 0.2   & 149 \\
28    & 1.8   & 0.1   & 255   & 1.87  & 0.05  & 271   & 2.35  & 0.03  & 58    & 6.6   & 0.1   & 929   && 11.3  & 0.1   & 52 \\
29    & 30.6  & 0.4   & 1765  & 27.3  & 0.3   & 1727  & 36.5  & 0.2   & 506   & 134.2 & 0.6   & 7739  && 875   & 1     & 8125 \\
30    & 6.2   & 0.1   & 990   & 6.2   & 0.1   & 961   & 19.1  & 0.1   & 662   & 43.8  & 0.2   & 7004  && 47.9  & 0.2   & 230 \\
31    & 13.9  & 0.3   & 882   & 9.4   & 0.2   & 472   & 9.2   & 0.1   & 198   & 50.5  & 0.4   & 3194  &&    284   &   0.7    & 295  \\
32    & 1.5   & 0.1   & 144   & 1.4   & 0.1   & 207   & 14.9  & 0.1   & 242   & 43.4  & 0.2   & 4222  && 10.2  & 0.1   & 205 \\
33    & 4.7   & 0.1   & 618   & 4.5   & 0.1   & 636   & 8.1   & 0.1   & 230   & 26.8  & 0.2   & 3521  && 114.1 & 0.3   & 1123 \\
34    & 4.8   & 0.1   & 532   & 4.4   & 0.1   & 642   & 20.0  & 0.1   & 446   & 55.1  & 0.3   & 6146  && 89.7  & 0.3   & 1290 \\ 
35    & 16.0  & 0.4   & 1986  & 13.6  & 0.3   & 2434  & 100.0 & 0.4   & 3308  & 271   & 1     & 33553 && 145   & 1     & 2379 \\
\hline
36    & 170   & 2     & 3115  & 111.6 & 1.7   & 2272  & 118.3 & 1.1   & 519   & 388   & 3     & 7085  && 3152  & 7     & 2864 \\
37    & 792   & 6     & 18910 & 683.9 & 5.0   & 21219 & 564   & 3     & 4330  & 463   & 6     & 11044 && 1705  & 12    & 4027 \\
38    & 11.2  & 0.3   & 424   & 8.1   & 0.2   & 419   & 7.1   & 0.1   & 44    & 18.0  & 0.4   & 680   && 105   & 1     & 599 \\
39    & 147   & 2     & 4701  & 100.5 & 1.7   & 4173  & 104   & 1     & 765   & 161   & 3     & 5123  && 347   & 7     & 1370 \\
\bottomrule
\end{tabular}
\end{table*}
\end{landscape}
\twocolumn

\end{document}